\documentclass[universe,review,accept,moreauthors,pdftex]{Definitions/mdpi} 

\newcommand{\fullcite}[1]{\citeauthor{#1} (\citeyear{#1}) \cite{#1}}
\newcommand{\fullcitep}[1]{(\citeauthor{#1} \citeyear{#1} \cite{#1})}
\NewDocumentCommand{\combocite}{ m m }{%
  \citeauthor{#1} (\citeyear{#1}), %
  \citeauthor{#2} (\citeyear{#2}) %
  [\citenum{#1}, \citenum{#2}]%
}
\NewDocumentCommand{\combocitep}{ m m }{%
  (\citeauthor{#1} \citeyear{#1}, %
  \citeauthor{#2} \citeyear{#2} %
  [\citenum{#1}, \citenum{#2}])%
}
\NewDocumentCommand{\combocites}{ m m m }{%
  \citeauthor{#1} (\citeyear{#1}), %
  \citeauthor{#2} (\citeyear{#2}), %
  \citeauthor{#3} (\citeyear{#3})
  [\citenum{#1}, \citenum{#2}, \citenum{#3}]%
}

\firstpage{1} 
\pubvolume{1}
\issuenum{1}
\articlenumber{0}
\pubyear{2024}
\copyrightyear{2024}
\datereceived{ } 
\daterevised{ } 
\dateaccepted{ } 
\datepublished{ } 
\hreflink{https://doi.org/} 

\Title{Selected Results on Variable Stars Observed by TESS}

\TitleCitation{Variable stars observed by TESS}

\Author{Zs\'ofia Bogn\'ar $^{1,\dagger,\ddagger}$\orcidA{} and \'Ad\'am S\'odor $^{1,\ddagger}$}

\AuthorNames{Zs\'ofia Bogn\'ar and \'Ad\'am S\'odor}

\AuthorCitation{Bogn\'ar, Zs.; S\'odor, \'A.}

\address[1]{%
$^{1}$ \quad HUN-REN Research Centre for Astronomy and Earth Sciences, Konkoly Observatory, MTA Centre of Excellence, Konkoly Thege Mikl\'os \'ut 15-17., H-1121 Budapest, Hungary; bognar.zsofia@csfk.org; sodor.adam@csfk.org}

\corres{Correspondence: bognar.zsofia@csfk.org}

\firstnote{These authors contributed equally to this work.}  
\secondnote{}

\abstract{As we enter the final year of the second extended mission of the Transiting Exoplanet Survey Satellite (TESS), it is time to reflect on what the TESS mission has contributed to the advancement of astronomy. Thousands of papers based on TESS data have already been published, making it a challenge to select the ones we mention or summarise in this review. As the title suggests, this paper focuses on variable stars, that is, phenomena that causes a star's brightness to change. {We discuss all the major classes of extrinsic and intrinsic variables, from planetary transits to pulsating stars, excluding only the longest-period ones, which are not well suited for the typical time spans of TESS time-series observations. TESS has provided significant and interesting data and results for all these variable types.} We hope that this selection successfully demonstrates the diverse applicability of TESS in variable star research.}

\begin{document}




\section{Introduction}

The need for nearly uninterrupted, high-quality time-series data on variable stars arose a long time ago and resulted in the formation of e.g. the Delta~Scuti Network \fullcitep{1997DSSN...11...37Z}), and the Whole Earth Telescope (WET; \fullcite{1990ApJ...361..309N}), the latter mainly aiming at the observations of compact pulsators. However, the most important source of extended data sets are photometric space telescopes, which can provide high-quality data covering a long, almost uninterrupted time base, and often observing a large portion of the sky, measuring a huge number of targets simultaneously. {Such data sets are well suited for the detection of planets orbiting stars other than the Sun, known as exoplanets. A commonly employed technique is the transit method, which detects shallow, periodic decreases in stellar brightness. These occur when an exoplanet, under favourable geometric alignment, partially obscures the host star’s light during its orbital motion.}

A series of photometric space telescope programmes helped and are helping us to reach these goals. Such missions are the Microvariability and Oscillations of Stars (MOST; \fullcite{2003PASP..115.1023W}), the Convection, Rotation and Planetary Transits (CoRoT; \fullcite{2006ESASP1306...33B}), the \textit{Kepler} (\combocite{2010Sci...327..977B}{2010PASP..122..131G}), the \textit{Kepler~2} (K2; \fullcite{2014PASP..126..398H}), and the Transiting Exoplanet Survey Satellite (TESS) space telescopes \fullcitep{2014SPIE.9143E..20R}. As most of these missions aim at the detection of different types of exoplanets, their observing characteristics allow us to perform the investigations of variable stars, too. This is also true for the forthcoming photometric space mission, the PLAnetary Transits and Oscillations of stars (PLATO; \fullcite{2025ExA....59...26R}), which is planned to be launched in 2026.   

{The main goal of this review is to summarise the most important scientific results on the variable stars observed in the framework of the TESS mission \fullcitep{2014SPIE.9143E..20R}. Admittedly, aiming at objectivity as best as the authors could, the summary of such a wast body of a continuously growing set of scientific results would inevitably be somewhat subjective. Whenever available, we focus on comprehensive studies of larger samples of variable stars of the particular classes, as exhaustively touching upon the thousands of individual case studies already published from TESS data is clearly impossible within the scope of this work. We discuss all the major classes of extrinsic and intrinsic variable stars, excluding only the longest-period ones, namely RV~Tauri, Mira and semiregular variables, which are not well suited for the typical time spans of TESS time-series observations (see details in Sect.~\ref{sect:tessintrod}), even though several studies were also published on these types of variables relying at least partially on TESS data.}



\section{Introducing the TESS space telescope}
\label{sect:tessintrod}

TESS is a NASA space telescope launched on 18 April 2018, as part of the Explorer programme. Its primary objective is to detect exoplanets orbiting bright, nearby stars using the transit method.

\begin{figure}[]
\includegraphics[width=1.0\textwidth]{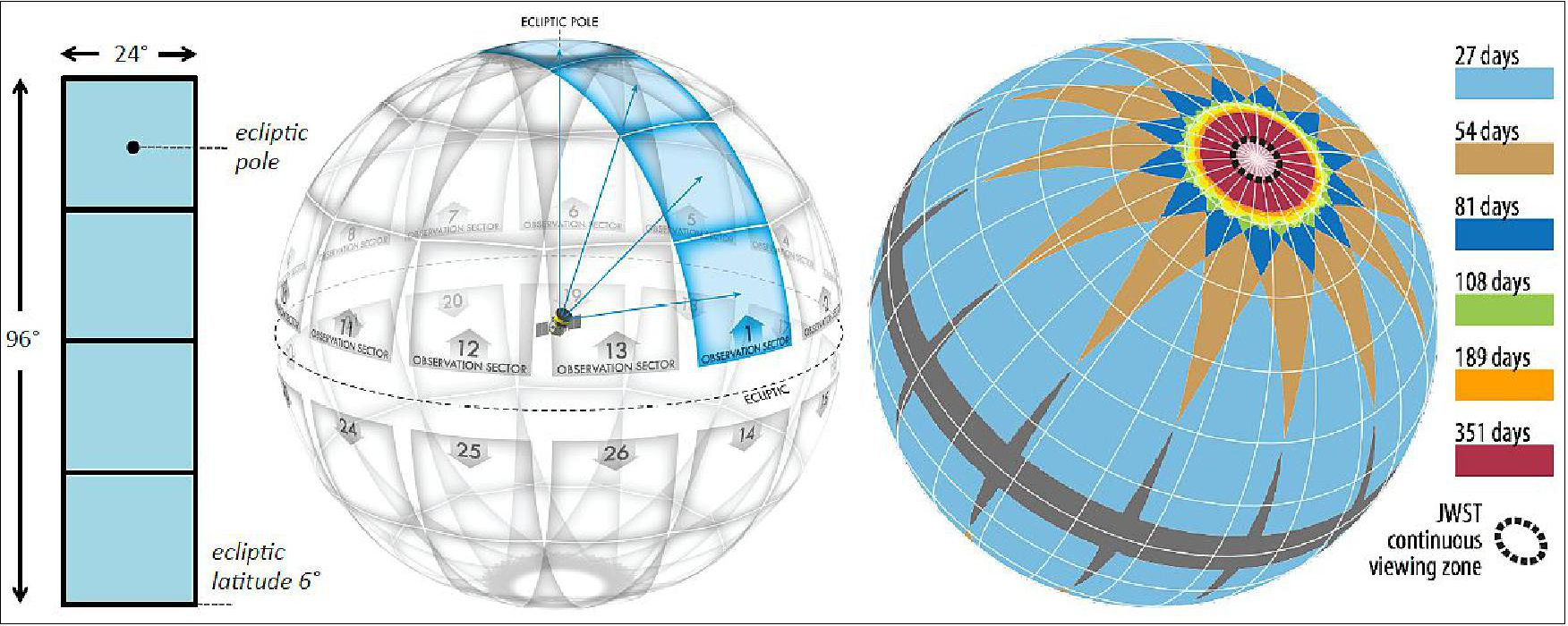}
\caption{Observational strategy of the TESS space telescope. Credit: TESS Team. \label{fig:sectors}}
\end{figure}

TESS follows a unique observational pattern that divides the sky into sectors of $24^{\circ} \times 96^{\circ}$, each of which is continuously observed for approximately 27 days (see Fig.~\ref{fig:sectors}). The spacecraft uses four wide-field cameras to cover almost 85\,--\,90\% of the sky over a two-year period.

The mission was designed to survey the sky in one-year cycles, starting with the southern ecliptic hemisphere (Cycle 1) and then moving to the northern one (Cycle 2). TESS orbits Earth in a highly elliptical, 13.7-day orbit at 2:1 orbital resonance with the Moon, optimised for continuous, stable observations with minimal interruptions.

Note that overlap between the different sectors occurs for the highest northern and southern ecliptic latitudes, which means that there are TESS targets that can be observed for more than 27 days in multiple consecutive sectors. Stars close to the ecliptic poles fall into the continuous viewing zone (CVZ), which is observed for almost a whole year.

\unskip


TESS collects data in different time cadences, depending on the scientific target: (1) Full-Frame Images (FFI): Initially taken every 30 minutes in the primary mission (Cycles 1 and 2, 2018\,--\,2020), it was then improved to a 10-minute cadence in Cycles 3 and 4 in the extended mission and was further improved to a 200-second cadence from Cycle 5 onwards. FFI data are available on each target in the field of view (FoV). (2) Short cadence mode: Selected stars were observed at a 2-minute cadence, allowing detailed analysis of their brightness variations. (3) Ultra-short cadence mode: Introduced in the first extended mission (Cycle 3, 2020), allowing 20-second exposures to study rapidly varying stars, such as compact pulsators.




The most important previous space mission for variable star studies was undoubtedly \textit{Kepler}, thanks to its high precision, {large field of view,} and about 4-year-long uninterrupted observation of the same set of targets with 30-min cadence. TESS has several advantages over \textit{Kepler}, but there are some disadvantages as well. The most important advantage of TESS over \textit{Kepler} is the significantly larger sky coverage: almost the full sky is observed, therefore the target sample size is much larger. The sample sets are much brighter on average and distributed much more evenly in the sky, making ground-based follow-up observations, especially spectroscopy, much more feasible. However, the large FoV and almost complete sky coverage also come with some compromises. The continuous time coverage is generally much shorter than that of \textit{Kepler}. Most of the targets are only observed separate, nonconsecutive 27-day-long sectors, generally once every two years, which makes the period analysis less precise and more susceptible to aliasing problems. In addition, the spatial resolution of the $\approx$\,21 arc-second pixel size of TESS is significantly worse than that of \textit{Kepler}, making contamination problems much more serious, especially in the crowded area of the Milky Way and other nearby galaxies, such as the Magellanic Clouds.

The main scientific contributions of TESS to various fields of astrophysics fall into two broad categories:

\begin{enumerate}

\item \textit{Exoplanet discovery}

The primary mission of TESS is to discover exoplanets around the brightest stars in the sky. Unlike its predecessor, the \textit{Kepler} Space Telescope, which mainly focused on a small, distant FoV, TESS scans a much larger portion of the sky, enabling the detection of relatively small planets around nearby stars. This makes verification follow-up observations easier and provides prime targets for future observations by the James Webb Space Telescope (JWST; \fullcite{2023PASP..135f8001G}) and other ground-based telescopes.

\item \textit{Stellar variability and asteroseismology}

Beyond exoplanet discovery, TESS also contributes significantly to stellar astrophysics by observing variable stars. The mission enables researchers to analyse stellar oscillations through asteroseismology, which helps to determine the internal structures of stars, and also to study stellar flares, binary systems, and other astrophysical phenomena.


\end{enumerate}

Following the success of its two-year primary mission (Cycles 1 and 2, 2018\,--\,2020), TESS was approved for an extended mission (Cycles 3 and 4, 2020\,--\,2022) and then for a second extended mission (Cycles 5 and 6, 2022\,--\,2025). A third extended mission is expected between 2025\,--\,2028. The extended missions introduced higher-cadence imaging, revisiting previously observed sectors, thus increasing the amount of data for long-term studies and improving sensitivity to different types of variable stars and exoplanets.

In the following sections, we present selected results considering variable star studies on stars showing extrinsic (Section~3) or intrinsic (Section~4) light variations. Note that the selection of variable star types and the presented scientific results are subjective to some extent, as we have no room to discuss all known types of variables and cite all main scientific results within the framework of this paper. The article ends with a discussion (Section~5).


\section{Stars showing extrinsic light variations}

Extrinsic light variations are stellar brightness variations observed specifically from our vantage point on Earth, which are caused by geometric effects, and do not originate from intrinsic variations of the stellar radiation. In this category, we review TESS results on eclipsing binaries, transiting exoplanets, and rotational light variations caused by spots on the stellar surface darker or brighter than their environment.

\subsection{Eclipsing binaries} 

Binary stars are pairs of gravitationally bound stars orbiting each other. {If the orbital plane is favourably aligned with the observer’s line of sight, periodic eclipses can be observed. These occur when one component passes in front of the other, partially or completely blocking its light.} The morphology of the light curve reveals a large amount of information on the binary components and on the geometry of their mutual orbit, such as stellar radii, surface temperatures, luminosities, semimajor axis, eccentricity, argument of periastron, and orbital inclination. If combined with radial velocity (RV), stellar masses and densities can also be measured. Binary and multiple systems are ubiquitous, they are present in every stellar population, and a significant fraction -- roughly half -- of them are members of a multiple system. Because of all this, binary stars are fundamental objects in stellar astrophysics. Binary stars sample star formation, stellar evolution, they are progenitors of Type Ia supernovae, they can produce nova eruptions, and are building blocks of higher-order multiple stellar systems. A large part of our present understanding of stellar formation and evolution relies on studying eclipsing binary stars (EBs).

Although they have been studied for more than two centuries \fullcitep{Goodricke}, and numerous aspects of EB systems are well understood today, many open questions remain. TESS observations helped and are expected to help in the future advance of our knowledge of these objects significantly. Since the physical properties inferred from the light- and radial-velocity-curves are independent of stellar evolution models, EBs are important targets to test stellar astrophysics, including stellar evolution, asteroseismology and gravitational theories, and to measure distances. Some of the most important open questions in the field of binary star studies are, e.g., the physical mechanisms determine multiplicity rates during star formation, and whether or not these are universal or environment-dependent; the physical background of the mass ratio distribution; stellar evolution in close binaries; the formation and evolution of exoplanets and exoplanetary orbits in binary star systems.

\begin{figure}[]
\includegraphics[width=1.0\textwidth]{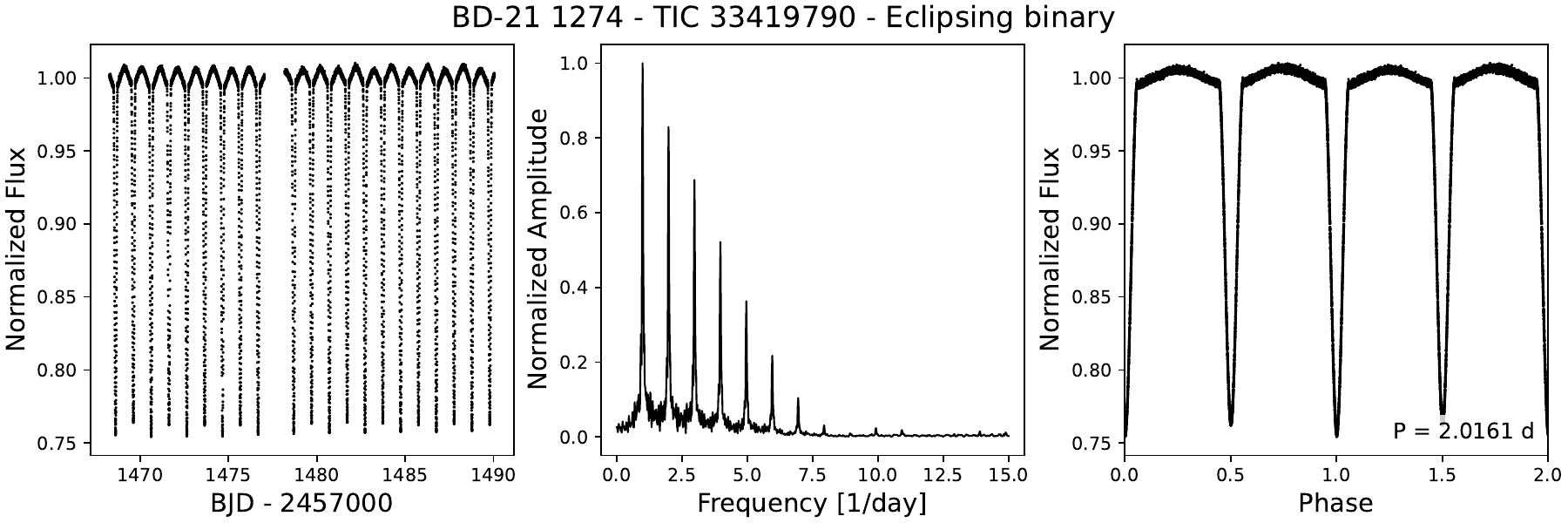}
\caption{A representative one sector long TESS light curve (left-hand panel), Fourier amplitude spectrum (middle panel) and light curve phased with the orbital period (right-hand panel) of an eclipsing binary, TIC~33419790 (BD-21\,1274) {\fullcitep{2022ApJS..258...16P}. Data was downloaded from the MAST Portal (https://mast.stsci.edu/portal/Mashup/Clients/Mast/Portal.html) and was processed by the authors. The light curve in the first panel shows the alternating primary and secondary minima of the eclipses. The spectrum in the middle panel is indicative of a highly non-sinusoidal, monoperiodic signal, with one fundamental frequency and many of it's harmonics. The photometric data folded with the orbital period in the third panel demonstrates the strictly monoperiodic, repetitive nature of the light variation.}}\label{fig:EB}
\end{figure}

Since EBs have a large degree of freedom in the physical parameter space, studying them comprehensively requires large homogeneous or carefully selected samples. Therefore, extensive photometric time series surveys, especially space measurements, are indispensable in these studies. TESS provides important advantages in observing the light curves of EBs in comparison to previous space missions. Many bright targets are observed across the whole sky, making spectroscopic radial-velocity follow-up observations easier, compared to photometric EB time series data from earlier space missions data. However, most of the targets are observed for just one 27-day-long sector continuously, which is appropriate only for short-period binaries. A representative TESS light curve of an eclipsing binary, TIC~33419790 (BD-21\,1274) is shown in Fig.~\ref{fig:EB}.

\unskip

\fullcite{2022ApJS..258...16P} collected 4584 EBs from the first two years of 2-min cadence TESS observations, some of which were already known before, but with many new discoveries. Due to the complicated target selection process, this sample is admittedly not homogeneous but still forms a very strong basis for future studies. \fullcite{2022ApJS..258...16P} determined the ephemerides for the whole sample. They present a wide variety of statistical analyses of the sample and propose that instead of keeping to the widely used discrete classes, we should rather use a binary star morphology classification in the future. Finally, \fullcite{2022ApJS..258...16P} announce an ongoing project to extract some 250 thousand EB light curves from the TESS FFIs, down to the 15th magnitude, using automated machine learning (ML) techniques.

\fullcite{2021A&A...652A.120I} found 3155 O-, B- or A-type EB candidates in the TESS light curve databases, focussing specifically on this least numerous group of stars. They developed a method capable of identifying periodic eclipses in the light curves even when relatively strong pulsational light variations contaminate the data. Their EB sample is largely disjunct from that of \fullcite{2022ApJS..258...16P}.

The unprecedentedly large EB sample of TESS allows interesting statistical verification of theories in ways that were not possible earlier. \fullcite{2024A&A...691A.242I} analysed the statistical properties of circularisation in a large sample of 14\,000 intermediate-to-high mass (O to F spectral class) eclipsing binaries observed by TESS in its first four years of operation. They explored the dependence of orbital circularisation on stellar properties and orbital parameters to improve our understanding of the physical processes affecting these systems. Their analysis revealed the dependence of orbital circularisation on stellar temperatures, also seen in other studies, and confirmed previous findings that additional dissipation is needed as compared to the predictions of turbulent viscosity and nonresonant radiative damping. They speculate that pulsations may play a role in the circularisation of close binaries.

Several interesting survey projects are going on that are looking deeper for EB light curves in the TESS data, such as \fullcite{2025arXiv250605631K}, where preliminary results indicate the identification of 10\,001 uniformly-vetted and -validated eclipsing binary stars in the first 82 sectors of TESS FFIs using ML techniques. These findings rely on the identification of more than 1.2 million EB candidates.

A particularly important EB discovery was made by \fullcite{2021AJ....161..162P} by using TESS data, who found the first known sextuply eclipsing sextuple star system, TIC~168789840. This multiple system consists of three eclipsing binaries in a hierarchical structure of an inner quadruple system with an outer binary subsystem. The highest level, widest separation subsystems were resolved by follow-up speckle interferometry. Their estimated orbital period is ~2\,kyr. The intermediate-level orbital period is ~3.7\,yr, while the innermost binaries in the quadruple member have orbital periods of $\sim$1\,d, and the outer standalone binary member have $\sim$8\,d orbital period. The three binaries are very similar to each other in terms of mass, radius and effective surface temperatures, therefore, this object can provide important new insight into the formation, dynamics, and evolution of multiple star systems. The inference of the widest and the intermediate orbital planes requires further observations.

TESS data are also the source of another exciting discovery in this domain. These are the Tidally Tilted Pulsators (TTPs), which are pulsating stars in close binaries, some of them are are also eclipsing. The pulsating component is tidally distorted to ellipsoidal shape. Due to the strong tidal forces, the axis of the pulsation can be tilted towards the companion star (see e.g. \fullcite{2020NatAs...4..684H}, \fullcite{2024ApJ...975..121J} and references therein). This means that during the orbital cycle, the pulsation can be observed from different latitudinal aspect angles -- almost 0\,--\,360$^{\circ}$ in high-inclination, eclipsing binaries. An interesting open question is why, in some frequency rich TTPs, are some pulsation modes tilted towards the companion, some remain untilted, while some might be tilted to an axis perpendicular to both. These are the triaxial pulsators, such as TIC~435850195 \fullcitep{2024ApJ...975..121J}. The tilted pulsations can help in mode identification, which, in turn, supports asteroseismic modelling. Therefore, TTPs are expected to be powerful tools for co-validating binary and asteroseismic modelling results on physical properties of the stellar members of these systems.

\subsection{Transiting exoplanets} 

The core design purpose of TESS is to discover exoplanets around dwarf stars in the solar neighbourhood, relying on the transit method. This method has proven to be the most productive one so far, with about three quarters of exoplanet discoveries based on this approach\footnote{\label{footnote1}https://exoplanetarchive.ipac.caltech.edu/docs/counts\_detail.html}. Therefore, exoplanetary results with TESS data have a huge and ever-growing literature. Exoplanet science is also a very captivating research field for the broader public, capable of raising interest and gaining support towards astronomical research in general. For these reasons, we include this field in our review, even though we are fully aware that we cannot aspire to completeness. Keeping this in mind, we briefly review the current status of this research domain related to TESS in this section.

The most fascinating goal of exoplanet research is arguably moving towards the possible discovery of extraterrestrial life in the universe. Therefore, the most interesting exoplanets are the Eart-like ones; rocky exoplanets about the size and mass of Earth, orbiting a star preferably similar to our own Sun, in the habitable zone (HZ), that is, at the distance where liquid water can be sustained on the surface of the planet. The difficulty of discovering transits of these kinds of exoplanets is the small fraction of light obscured by the transiting exoplanet because of the large radius difference between the planet and host star and, due to the long orbital period, the scarcity of the transit events.

Due to these difficulties, the transit method strongly distorts the sample of found exoplanets towards short-period, gas giant planets close to their host star, since all three factors increase the chance of detection. A shorter orbital period means more transit events in the given observing interval, a larger radius of the exoplanet relative to its host star causes deeper, higher S/N transits, and a closer orbit permits a wider viewing angle from which the transits are observable. Therefore, the most populous class of known, confirmed exoplanets are the hot Jupiters and hot Neptunes, which are gas giants orbiting close to their host stars, with sizes similar to that of Jupiter and Neptune, respectively. At the same time, very few small-radius, Earth-like, rocky planets are known in the HZs of the host stars, with orbital periods on the order of dozens or even hundreds of days.

\begin{figure}[]
\includegraphics[width=1.0\textwidth]{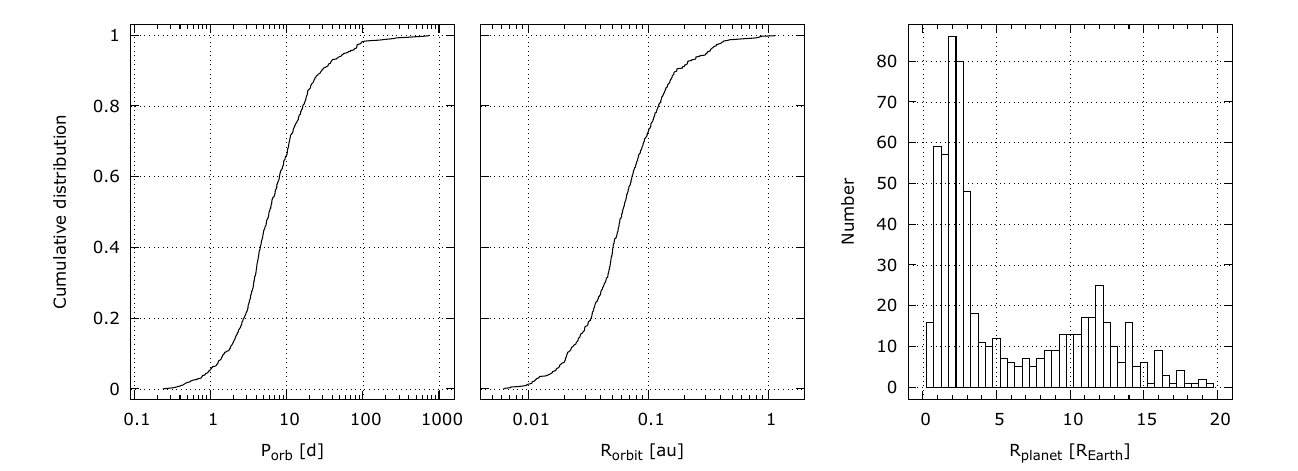}
\caption{\label{fig:exoplanetstats}
Cumulative distributions of orbital period, orbital radius and planet radius of the 638 confirmed TESS exoplanets. Note that of the 638 objects, only 600 have computed $R_{\textrm{orbit}}$ and 634 have computed $R_{\textrm{planet}}$ parameters available.}
\end{figure}

\begin{figure}[]
\includegraphics[width=1.0\textwidth]{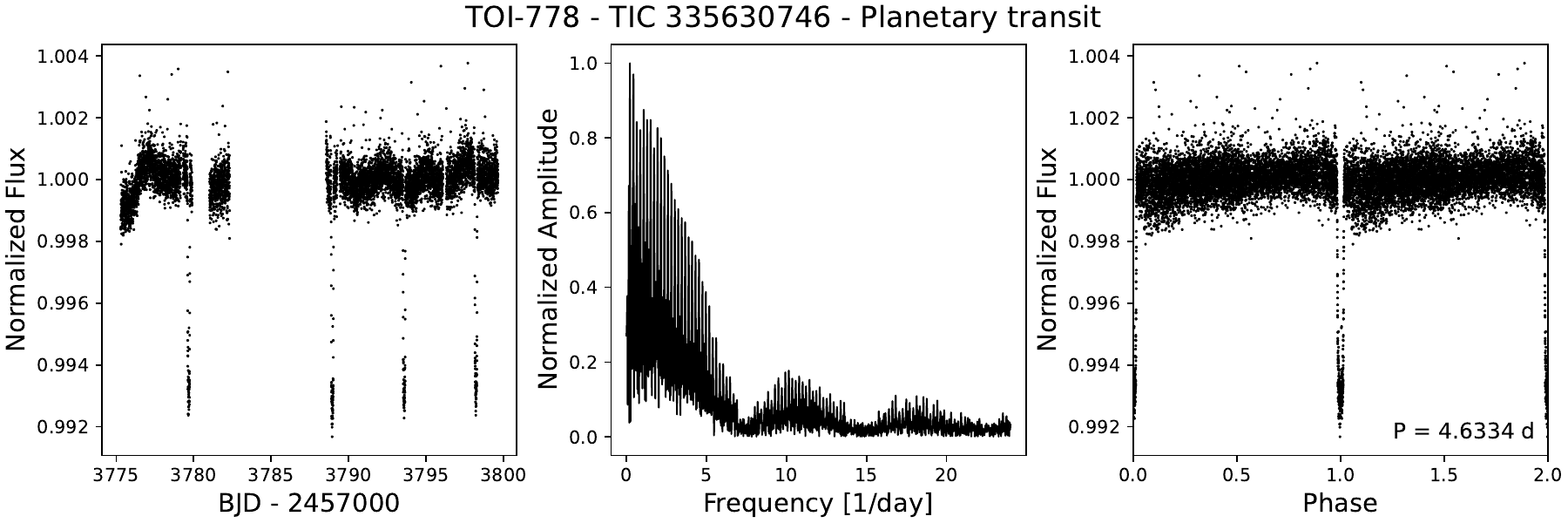}
\caption{{Representative TESS photometric data of a transiting exoplanet around TIC~335630746 (TOI-778) \fullcitep{2023AJ....165..207C}. The structure and data source of the figure is the same as of Fig.~\ref{fig:EB}.}\label{fig:transit}}
\end{figure}

At the time of writing this part (01 July 2025), 638 confirmed and 7655 candidate transiting exoplanets were discovered in the TESS data\footnotemark[1]. We plotted some basic statistics of the 638 confirmed TESS exoplanets in Fig.~\ref{fig:exoplanetstats}, such as distribution of orbital periods, orbital radii, and planetary masses. From the first panel of Fig.~\ref{fig:exoplanetstats}, we can see that 66\,\% of the sample has an orbital period shorter than 10\,d, while only 2\,\% has an orbital period above 100\,d. The radius distribution shown in the third panel of Fig.~\ref{fig:exoplanetstats} is bimodal, with a significant radius gap at around 7 $R_{\textrm{Earth}}$, separating Neptune- and Jupiter-size exoplanets. A representative TESS light curve of a transiting exoplanet host star, TIC~335630746 (TOI-778) is shown in Fig.~\ref{fig:transit}.

\unskip

\unskip

{The main advantage of TESS over \textit{Kepler} in exoplanetary research is that the targets are generally brighter, therefore, they are much better suited for follow-up studies to confirm and further investigate planetary candidates by spectroscopic and imaging observations. These efforts are coordinated by the TESS Follow-up Observing Program (TFOP) Working Group (WG)\footnote{https://tess.mit.edu/followup/}. One of the many activities of this WG is the TESS-Keck Survey's (TKS) Mass Catalog, which is a surveying programme on the Keck I\,--\,HIRES telescope\,--\,spectrograph pair aimed specifically at obtaining a large, homogenous RV sample on exoplanet host stars discovered by TESS \fullcitep{2022AJ....163..297C}. The density, characterised by the mass\,--\,radius relation is one of the most important parameters of exoplanets, which may have some scatter from various sources. The intrinsic variability of the chemical composition of the planets is one such source, but the inhomogeneity of measurements with different instruments is another one. The homogeneity of the TKS ensures to eliminate the latter one, so the intrinsic variations can be investigated with higher confidence. Some of the latest results of the TKS is summarised by \fullcite{2024ApJS..272...32P}. The authors highlight the confirmation of 32 new planets of varying sizes through RV measurements and
statistical validation. Their latest catalogue contains 5110 Keck/HIRES RVs and activity indicators.}

{The first habitable zone terrestrial planet discovered by TESS is TOI-700\,d \fullcitep{2020AJ....160..116G}. TOI-700 (TIC 150428135) is an inactive M2.5 dwarf star 31.1\,pc away from us, hosting 4 known planets, of which two, TOI-700\,d and the later discovered TOI-700\,e are Earth-sized, and orbit in the HZ of the host star \combocitep{2020AJ....160..116G}{2023ApJ...944L..35G}. Due to the brightness of the host star and the multi-planet system, TOI-700 is a valuable target for follow-up observations by current and near-future facilities. An interesting, and still unanswered question is whether these two, supposedly rocky planets could have maintained their atmospheres, given the magnetic activity of their host star, since the accompanying flare events can erode the atmospheres away.}

{Predicting transit times is important for scheduling follow-up observations in the future, which is based on precise measurement of times of transits in past observations. The study of transit timings can also reveal variations in transit times (TTV), which may indicate a plethora of dynamical phenomenon, such as multi-planetary interactions, planetary mass loss, tidal orbital decay etc. \fullcite{2022ApJS..259...62I} compiled a large database of transit times based on TESS and other observations. The database consists of 8667 transit-timing measurements for 382 systems. About 240 planets in the catalogue are hot Jupiters observed by TESS. Beyond the previously know case of WASP-12b \fullcitep{2016A&A...588L...6M}, the authors identified several candidate hot Jupiters showing hints of TTV.}

{\fullcite{2025NatAs...9.1007K} studied the dynamical environment, specifically the eccentricity distribution of all known transiting Earth proxies -- planets with similar sizes and stellar irradiation to Earth -- orbiting late-type dwarf stars. Their initial sample consisted of 68 \textit{Kepler} and 26 TESS targets, while after further filtering, 17 exoplanets remained in the final sample. The main conclusion of this study is that the Earth proxies are dynamically cool, that is, with one exception, all the targets show near circular orbit. The open question is whether the single one remaining eccentric orbit (e\,$\sim$\,0.4) represents a subgroup of highly eccentric orbits, or is purely just an outlier. Anyhow, the result suggests that the vast majority of Earth proxy planets have near circular orbits, which is favourable for climate stability.}

Identifying the signatures of transiting exoplanets in tens of millions of light curves requires a high degree of automation and sophistication. Therefore, this field of research employs various cutting-edge ML techniques, most of which rely on convolutional neural networks (CNNs). The fast development of this area is demonstrated by the fact that just during the weeks preceding the finalisation of this manuscript (July 2025), two new studies on two new ML-techniques in this domain were published in the literature. Among the latest developments are \fullcite{2025AJ....170...21H}, who propose a deep learning model architecture using long short-term memory (LSTM) networks and additive attention mechanisms for vetting transiting exoplanet candidates, with focus on \textit{Kepler} and TESS light curves. The authors conclude that their model can reach performance in identifying transiting exoplanet light curve features similar to more traditional CNNs, but with less complexity, that is, with fewer trainable parameters. This makes the cost of training higher, but the application of their model cheaper. Furthermore, \fullcite{2025AJ....170...73F} recently developed a deep learning CNN model, named {\tt DART-Vetter}, which is able to distinguish exoplanetary candidates from false positive signals. The advantage of their model over earlier models is simplicity and compactness, since {\tt DART-Vetter} processes only the light curves folded with the period of the relative signal. {These models, however, are only the latest ones, and the question is still open, whether or not they would perform better on future data than previous, more established models, such as Astronet \fullcitep{2018AJ....155...94S} or ExoMiner \fullcitep{2022ApJ...926..120V}.}


As the TESS mission is still going on, the observation baselines get extended, while the identification methods get more sophisticated, we can expect that the rate of discovery of exoplanet candidates will even increase. At the same time, the verification process of the discovered exoplanet candidates will take many more years, even after the conclusion of the mission. Therefore, we can expect many interesting results and large-sample comprehensive studies in this field in the future.

\subsection{Light variations by stellar rotation} 

Several physical mechanisms can lead to brightness variations in stars as a result of their rotation. One common phenomenon is the existence of surface spots (starspots), which are similar to the sunspots, but often much larger. These are cooler, darker regions on the stellar surface, caused by magnetic fields. As the star rotates, these spots move in and out of view, causing periodic dips in brightness. This results in rotationally modulated light variations.

{\fullcite{2024AJ....167..189C} analyzed individual TESS sectors to detect spot modulation caused by short-period (less than $\sim$12\,d) stellar rotation, using only parameters measured from light curves for a robust and unbiased method of evaluating detections. The authors used data from the first 26 sectors of TESS. They analyzed 432\,704 short, 2 minutes cadence single-sector light curves for FGKM dwarfs. They detected 16\,800 periods in individual sector light curves, covering 10\,909 distinct targets, and they also presented a catalogue of the median period for each target.}

\fullcite{2024ApJ...962...47C} used deep learning on year-long TESS FFI light curves to derive reliable rotation periods for more than 7000 main-sequence stars near the southern ecliptic pole, extending rotation period measurements beyond TESS's previous limit of 13.7 days. Their results confirm known features of stellar rotation, such as the intermediate-period gap and its link to reduced photometric variability. Rotation detectability correlates with temperature, metallicity, and spot activity. They show that deep learning can overcome TESS systematics, and their method holds promise for future missions such as the \textit{Roman} Space Telescope \fullcitep{2015arXiv150303757S}.

Besides cool starspots, we know hot spots, which are regions on the stellar surface that are hotter and brighter than their surroundings, often due to accretion from a companion star, or magnetic activity.

\begin{figure}[]
\includegraphics[width=1.0\textwidth]{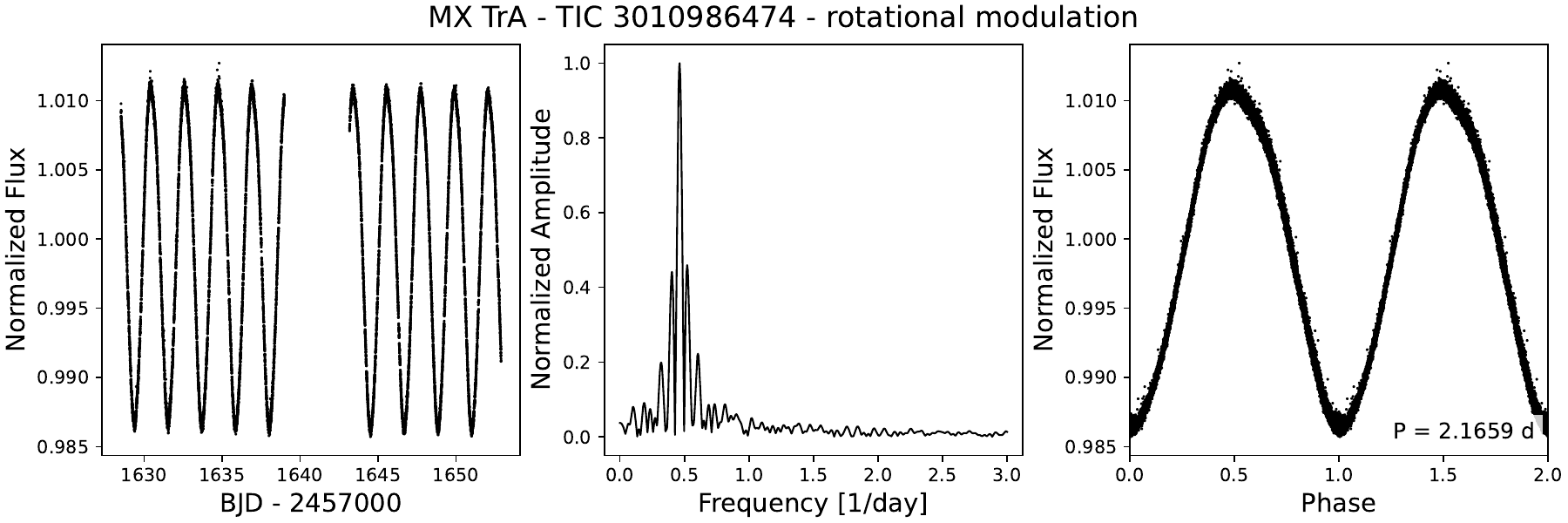}
\caption{{Representative TESS photometric data of a rotationally modulated chemically spotted star, TIC~3010986474 (MX~TrA) \fullcitep{2024Univ...10..341P}. The structure and data source of the figure is the same as of Fig.~\ref{fig:EB}.\label{fig:rotmod}}}
\end{figure}

Another source of light variations by stellar rotation is the chemical surface inhomogeneities, which are especially common in magnetic, chemically peculiar stars (e.g., Ap stars). Certain elements may be overabundant or depleted in specific surface regions, affecting the emitted light. As these regions rotate in and out of view, they cause brightness changes. An example of this was given by \fullcite{2024Univ...10..341P} who presented the modelling results of the TESS light curve of the Ap Si star MX~TrA. They were able to reproduce the shape of the observed light curve and its amplitude with an accuracy better than 0.001 mag. They found that the inhomogeneous surface distribution could account for the elements Si, Fe, Cr, and He. {An example TESS light curve of a rotationally modulated chemically spotted star, TIC~3010986474 (MX~TrA) is shown in Fig.~\ref{fig:rotmod}.}

\unskip

Note that starspots are often caused by magnetic activity, which results in intrinsic variation of the stellar brightness through changing spot coverage and also flare events, which leads us nicely to our next section.



\section{Stars showing intrinsic light variations}

Intrinsic light variations mean that the star's luminosity varies because physical changes happen above or on the stellar surface. In the following sections, we introduce some of the TESS results concerning stars with intrinsic light variations.

\subsection{Eruptive variables} 

The General Catalogue of Variable Stars (GCVS5.1; \fullcite{2017ARep...61...80S}) lists 21 types of eruptive variables. This group encompasses pre-main sequence stars, along with S~Doradus, $\gamma$~Cassiopeiae, and Wolf-Rayet stars. In these objects, a sudden influx of energy occurs within the star or part of it, triggering the outburst that we observe. 

Due to the long time scales involved in some eruptive phenomena, such as those observed in FU~Orionis, Wolf-Rayet, and luminous blue variables, events that can extend to even decade-long intervals, TESS is obviously not capable of observing such behaviour. However, all the progenitor stars of these eruptive phenomena show variability on shorter time scales that can be investigated in TESS data.


\subsubsection{Flare stars}

\begin{figure}[]
\includegraphics[width=1.0\textwidth]{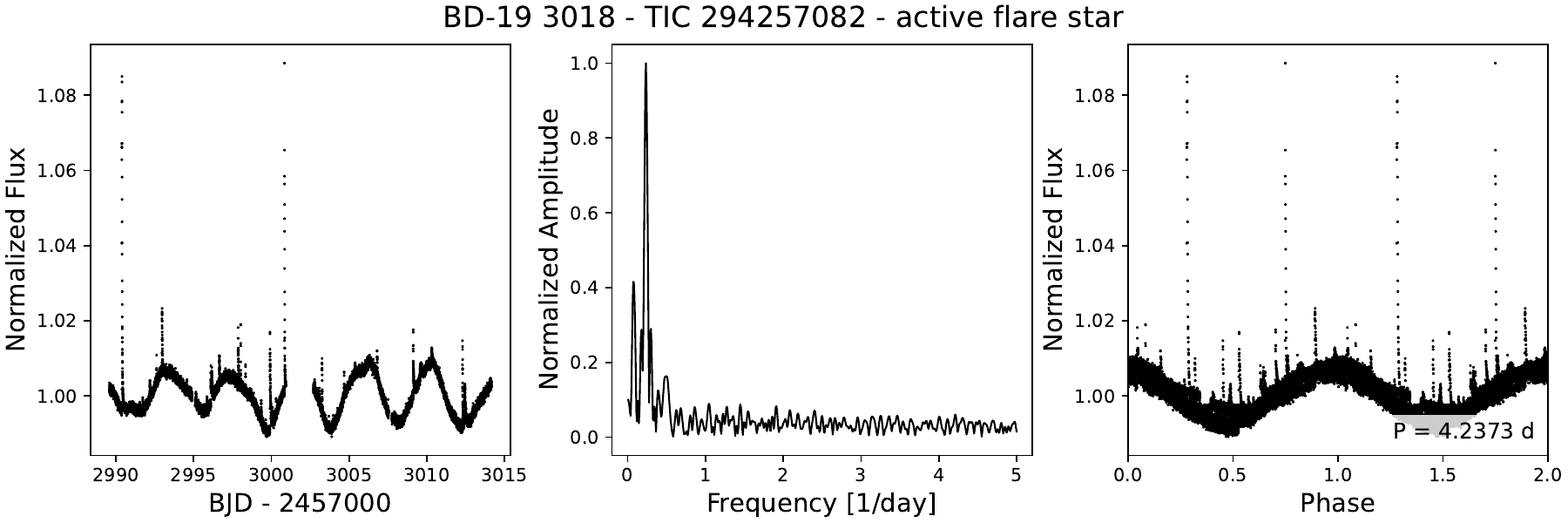}
\caption{{Representative TESS photometric data of the flare star, TIC~294257082 (BD-19~3018) \fullcitep{2020MNRAS.494.3596D}. The structure and data source of the figure is the same as of Fig.~\ref{fig:EB}. Note that rotational modulation due to starspots is also observable. We used the spot modulation period for the phased light curve in the right panel.\label{fig:flare}}}
\end{figure}

Flare stars are M- or K-type red dwarfs that produce sudden, intense flares because of strong magnetic activity. They also often show rotational spot modulation. A representative TESS light curve of a flare star, TIC~294257082 (BD-19~3018) is shown in Fig.~\ref{fig:flare}. Stellar flares play a crucial role in shaping exoplanetary environments. Although they can erode planetary atmospheres and reduce habitability, they may also contribute to the chemical processes necessary for life, particularly around low-mass stars.

\fullcite{2020AJ....159...60G} investigated stellar flares using data from the first two observation sectors of TESS. During these two months, TESS observed 24\,809 stars with a 2-minute cadence, and they were able to identify 1228 flaring stars, of which 673 are M dwarfs. In total, they detected 8695 flares. They found that 30\% of mid-to-late M dwarfs, 5\% of early M dwarfs, and less than 1\% of F-, G- and K-type stars exhibit flaring activity. Among flaring M dwarfs, they identified 531 early-type and 142 late-type M dwarfs. {The authors solidified past findings that fast rotating M dwarfs are the most likely to flare and that their flare amplitude is independent of the rotation period.}

\fullcite{2020AJ....159...60G} also examined their findings in the context of prebiotic chemistry, coronal mass ejections (CMEs), and ozone depletion. Stellar flares may provide sufficient ultraviolet energy to trigger biogenesis on exoplanets. They identified 14 stars, including 11 M dwarfs, that meet the criteria for the necessary flare rate and energy. However, flares associated with CMEs and stellar particle events (SPEs) could negatively impact planetary atmospheres, potentially depleting the ozone layer.

{\fullcite{2023A&A...669A..15Y} precisely detected 60\,810 flare
events on 13\,478 stars from sectors 1\,--\,30 of the TESS data and determined their parameters with the aim of determining the relationship between flare parameters and stellar parameters. Stellar parameters were taken from the \textit{Gaia} survey. The authors found that the fraction of flaring stars decreases as stellar temperatures increase in the region of 2500\,--\,6500\,K and stellar mass increases from 0.08 to 1.4\,\,$M_\textrm{Sun}$. The flare energies increase as the stellar temperature and the stellar mass decrease. The authors also confirmed that M-type stars produce flares more frequently than F-, G-, and K-type stars. They
found that the proportion of flaring stars increases from M0 to M5, and decreases from M5 to M7 spectral type.}

\fullcite{2025A&A...694A.161S} searched for flares in the first five years (sectors 1–69) of the TESS mission using 2-min cadence observations. Using these data, they created a high-purity catalogue of $\sim$120\,000 flare events on $\sim$14\,000 stars. With this data set, they found, for example, that compared with flares in solar-like stars, the plasma responsible for M-dwarf flare emission is denser, and flares probably originate from a deeper layer of the stellar atmosphere.

{TESS is an important data source in flare star studies, especially considering easier follow-up observations, but compared to \textit{Kepler} measurements, it does not represent a significant qualitative advancement.}

\subsubsection{Wolf–Rayet stars and luminous blue variables}

\begin{figure}[]
\includegraphics[width=1.0\textwidth]{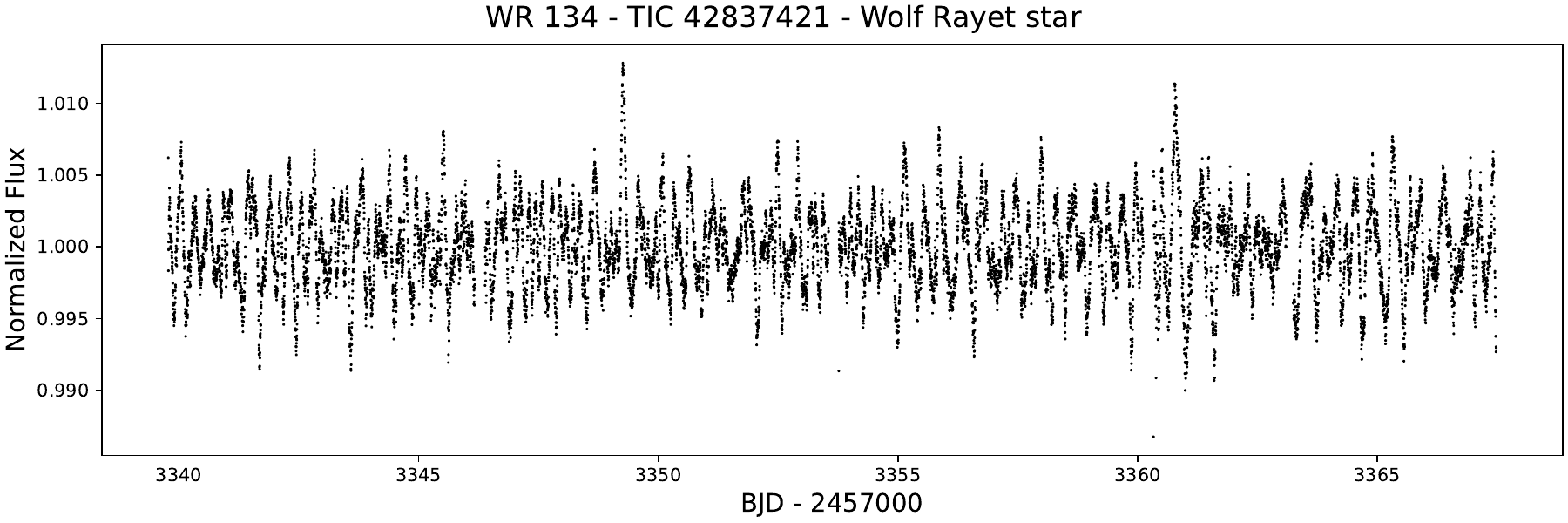}
\caption{A representative one sector long TESS light curve of the WR star, TIC~42837421 (WR~134) \fullcitep{2021MNRAS.502.5038N}. {The data source and processing are the same as of Fig.~\ref{fig:EB}. Note that there are no characterisctic periodicities in the data, therefore the Fourier amplitude spectrum and folded light curve is not shown here.}\label{fig:fuor}\label{fig:wr}}
\end{figure}

Wolf-Rayet (WR) stars are massive, evolved stars that are in a late stage of their stellar evolution. They are characterised by their strong, broad emission lines, extreme mass loss, and high surface temperatures. Luminous blue variables (LBVs) are massive, evolved stars that undergo extreme variability in brightness and mass loss. Their brightness fluctuates over months to decades, with variations of up to 2 magnitudes in optical wavelengths. They show unpredictable outbursts, where they loss large amounts of mass. We can observe high-frequency variability of WR and LBV stars \fullcitep{2021MNRAS.502.5038N}. However, their light curves display low-frequency stochastic variability ('red noise'), too. \fullcite{2021MNRAS.502.5038N}, using TESS observations, did not find coherent, isolated signals for LBVs but detected such signals in five WR stars. A representative TESS light curve of a WR star, TIC~42837421 (WR~134) is shown in Fig.~\ref{fig:wr}. 

\unskip

\subsection{Cataclysmic variables} 

Cataclysmic variables are interacting binary stars where the so-called primary star is a white dwarf, and the secondary star in most cases is on or close to the main sequence. They orbit so close to each other that matter is transferred from the Roche-lobe-filling secondary star to the primary star. 

{Variable stars belonging to this class are unique in the sense that cataclysmic events are hard or impossible to predict and often are not recurrent. Also, due to the immense brightness increase, the quiescent progenitor stars are often not detectable. We still include these variables in our review, since TESS is an important surveying instrument capable of discovering many such events. An important factor here is the monitored area of the sky, in which TESS has a real advantage over \textit{Kepler}.}


\subsubsection{Dwarf novae and novae}

\begin{figure}[]
\includegraphics[width=1.0\textwidth]{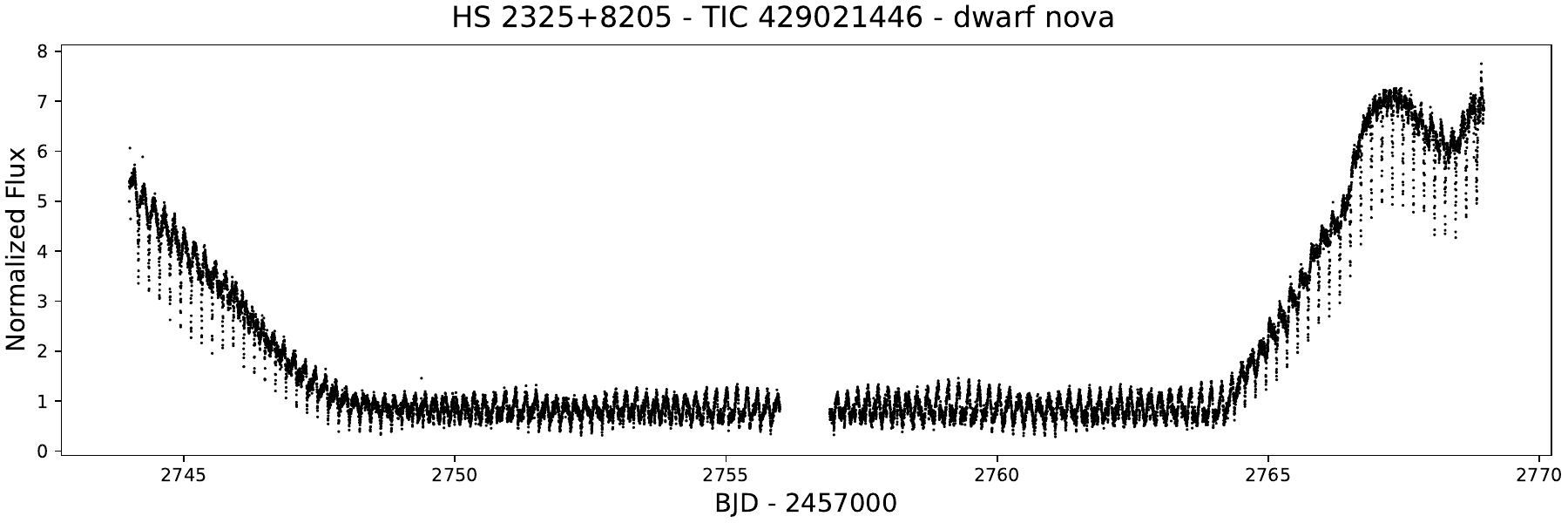}
\caption{{A representative one sector long TESS light curve of the dwarf nova, TIC~429021446 (HS~2325+8205) \fullcitep{2023MNRAS.518.3901S}. The data source and processing are the same as of Fig.~\ref{fig:EB}.} Note that this sector mostly covers a quiescent period, but the end of a previous outburst event, and the beginning of the next one can be observed. The quick periodic variations are due to eclipses of the member stars and the accretion disc around the erupting component.\label{fig:dwarfnova}}
\end{figure}

Dwarf novae show large (2\,--\,6 magnitudes) brightening from time to time. The physical cause of this is that the flow of inward material increases significantly, which increases the luminosity of the disc by heating. TESS also observed dwarf novae, for example the star AY~Psc, which is an eclipsing system. The TESS observations contributed to an improvement of its orbital period. Unfortunately, \fullcite{2023MNRAS.519..352B} did not observe the brightening phase of this system during the TESS observations. The situation is different for the dwarf nova HS~2325+8205, where the authors presented that the system has four different types of outbursts \fullcitep{2023MNRAS.518.3901S}. The authors derived an outburst period of about 13.83\,d, and, for the first time, they presented that there are quasiperiodic oscillations with a period of $\sim 2160$\,s in the long outburst of HS~2325+8205. An example TESS light curve of this dwarf nova is shown in Fig.~\ref{fig:dwarfnova}.

\unskip

Novae show even larger brightness increases than dwarf novae: generally 10\,--\,12 magnitudes. The cause of this brightness variation is that part of the outer layer of the accreting white dwarf gets pushed off, accompanied by a thermonuclear burst. From the observers' view, the brightening phase is very quick on a time scale of days, whereas the dimming phase can last up to about 100 days. TESS observed several novae, presented for example in two papers \combocitep{2022MNRAS.514.4718B}{2023MNRAS.519..352B}. A specific type of persistent modulation observed in many CVs is known as superhumps, which are variations with a period differing by a few percent from the binary's orbital period. These papers mostly focus on this superhump phenomenon. Superhumps are one of the most common variability phenomena detected in the light curves of cataclysmic variables. The TESS data demonstrate that the occurrence of superhumps in novalike and old novae subtypes is not an exception, but quite common.

\subsubsection{Supernovae}

Supernovae are stars that undergo a sudden, catastrophic outburst, increasing in brightness by 20 magnitudes or more before gradually fading. {The supernova nucleosynthesis is a major source of heavy elements in the universe.}


Based on their physical cause, light curve characteristics, and spectral features, supernovae are classified into two main types: I and II. The presentation of further details on the physical properties of novae and supernovae is beyond the scope of this paper. However, we also briefly summarise some TESS-related results on supernovae. TESS observations helped reveal that the unusual Type Ia supernova ASASSN-18tb is the only SN Ia showing H$\alpha$ from circumstellar medium interaction to be discovered in an early-type galaxy \fullcitep{2019MNRAS.487.2372V}. Furthermore, it was the first extragalactic transient to be studied with TESS data. Considering the core-collapse supernovae, twenty bright ones were investigated in the cited study \fullcitep{2021MNRAS.500.5639V}. They found an average pre-rise flux excess of approximately 2.5 per cent of the peak flux, occurring around one day before the estimated onset of the rising light curve, likely attributed to shock breakout.

\begin{figure}[]
\includegraphics[width=1.0\textwidth]{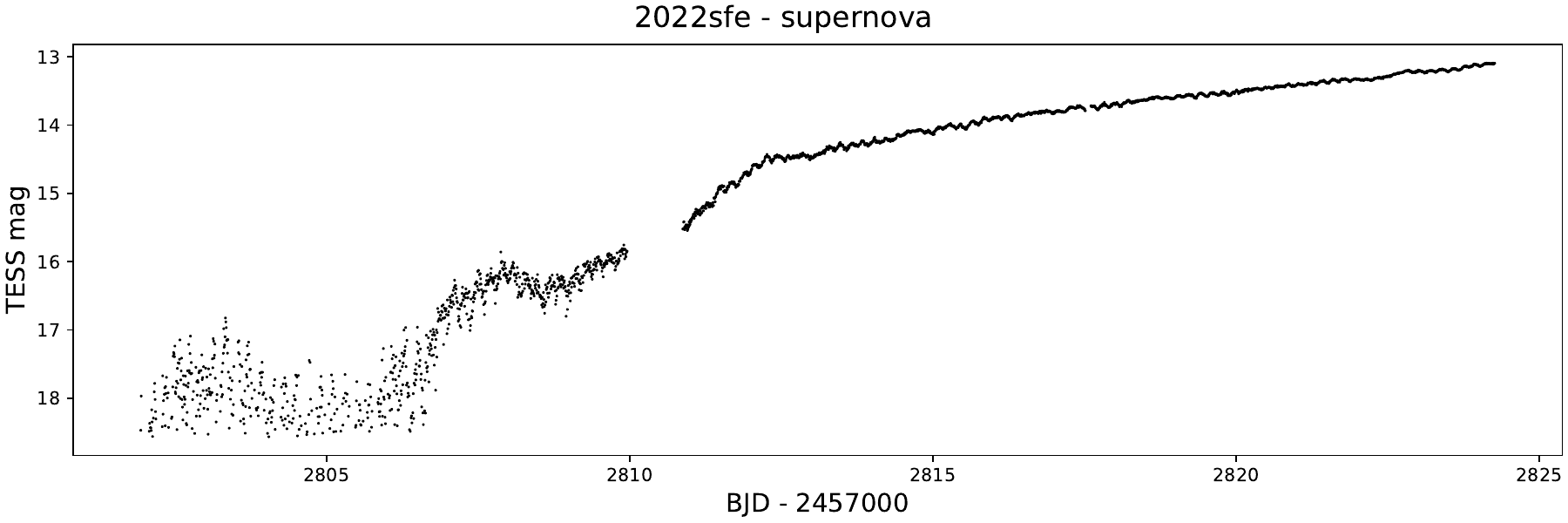}
\caption{One sector long TESS light curve of the outburst of the supernova SN2022sfe. The initial brightness is most probably below the sensitivity limit of TESS, but at some time around JD\,2\,459\,806, the object is positively identified, and the rest of the sector covers most of the brightening phase. {Note that there are no characterisctic periodicities in the data, therefore the Fourier amplitude spectrum and folded light curve is not shown here. The data was downloaded from the TessTransients webpage (https://tess.mit.edu/public/tesstransients/pages/sector55-all-transients.html).}\label{fig:sn}}
\end{figure}

Due to its design, TESS is not the perfect instrument to follow up supernovae because of the long duration of the decline, and the necessity of multicolour follow-up measurements. However, thanks to its survey characteristic, TESS is very useful in observing the initial brightening phase right from the beginning. One such example is shown in Fig.~\ref{fig:sn}.

\unskip

\subsection{Stellar pulsation}

Many stars show intrinsic light variations caused by stellar pulsations. The Hertzsprung-Russell diagram (HRD) is almost completely covered by the different types of pulsating variable stars. In the following subsections, we present the TESS results concerning some selected groups of variable stars, such as RR~Lyrae, Cepheid, $\delta$~Scuti, $\gamma$~Doradus, roAp, {$\beta$~Cep, Slowly Pulsating B (SPB) stars,} Solar-like, white dwarf, and hot subdwarf pulsators.

{In these cases, the period and amplitude of brightness variations are appropriate for discovering many new variable stars using TESS data, and for detecting pulsation modes for asteroseismic studies.}

\subsubsection{RR~Lyrae stars} 

RR~Lyrae stars are old, core He-burning stars with typical masses in the 0.5\,--\,0.8\,\,$M_\textrm{Sun}$ range, located at the intersection of the classical instability strip and the horizontal branch in the HRD. Their large-amplitude pulsations {(typically 0.2\,--\,1.6\,mag; \fullcite{2009AcA....59..137S})} are dominated by fundamental or first-overtone radial modes. However, recent ground- and space-based observations revealed additional non-radial pulsation modes in their light variations. Furthermore, periodic modulation of the shape and phase of the light curve is observed in many RR~Lyrae stars, which is called the Blazhko effect. The origin of the additional modes and the underlying physical mechanism of the Blazhko effect are the main open questions in RR~Lyrae research. Hopefully, the TESS observations will help us in our way towards finding answers to these questions.

\begin{figure}[]
\includegraphics[width=1.0\textwidth]{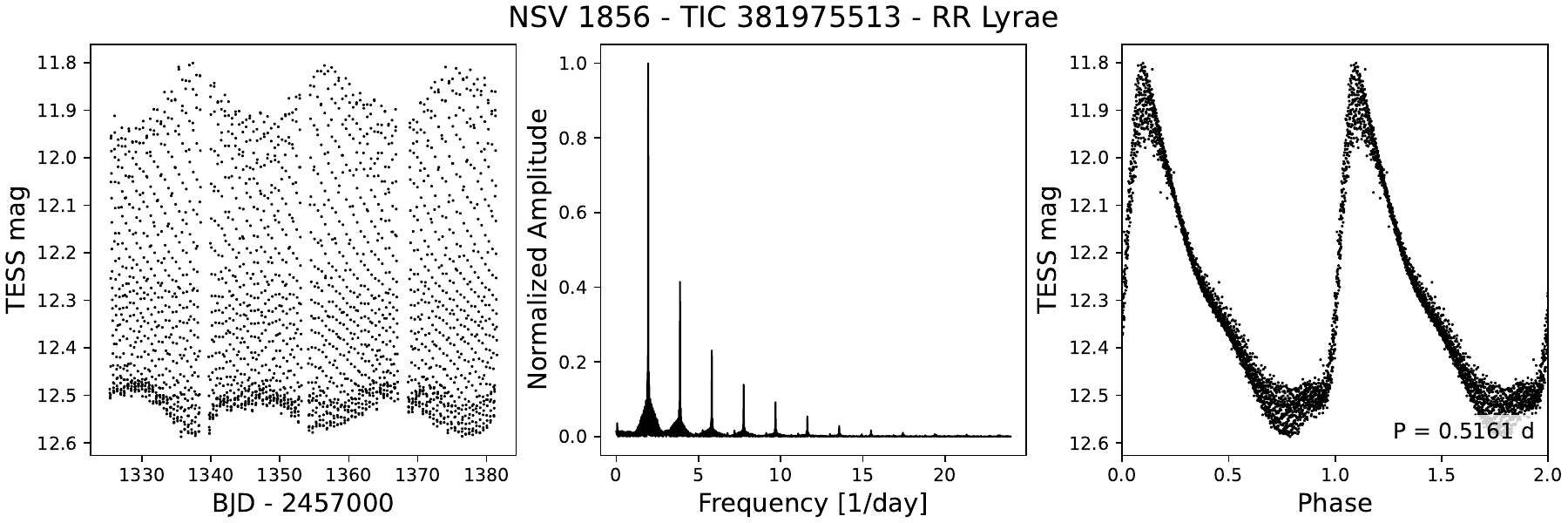}
\caption{{Representative TESS photometric data of an RR~Lyrae star, TIC~381975513 \fullcitep{2022ApJS..258....8M}. The structure of the figure is the same as of Fig.~\ref{fig:EB}.} This fundamental-mode RR~Lyrae shows the Blazhko effect, which is the modulation of the light curve shape, observable both on the left panel, as sinusoidal upper and lower envelopes, and in the right panel as an increased scatter around minimum and maximum light phases. We used the light curve data published by \fullcite{2022ApJS..258....8M}. \label{fig:rrl}}
\end{figure}

TESS data has several advantages, but also disadvantages over \textit{Kepler} and K2 in studying RR~Lyrae stars. The advantages are: higher cadence, as TESS observes later sectors with 10 minute resolution, which improves frequency resolution and decreases phase smearing; broader sky coverage enables selection from a larger sample and permits easier follow-up observations with smaller ground-based telescopes limited to brighter targets; combining accurate TESS data with homogeneous parallax and therefore absolute brightness and colour data from Gaia Early Data Release~3 \fullcitep{2021A&A...649A...1G} offers us the best photometric classification scheme for classical pulsators \fullcitep{2022ApJS..258....8M}. Some of the disadvantages are: less accuracy due to the smaller telescope aperture; a shorter observational duration of TESS limits the ability to detect long-term modulations, such as the Blazhko effect, which may require extended monitoring; lower angular resolution makes it harder to separate variations from neighbouring stars. A representative TESS light curve of a fundamental-mode RR~Lyrae star, TIC~381975513 is shown in Fig.~\ref{fig:rrl}.

\unskip

Investigating the potential of the TESS space telescope in RR~Lyrae studies, \fullcite{2022ApJS..258....8M} performed a comprehensive analysis of 126 known or candidate bright RR~Lyrae stars within Sectors 1 and 2. Among these targets, 118 turned out to be real pulsators. The key findings of their work are: Combining accurate TESS light curves with homogeneous Gaia parallaxes provides absolute brightness and colour data that offer an effective photometric classification scheme for classical pulsators, especially distinguishing between RR~Lyrae and anomalous Cepheids. Although the short time span of the TESS sectors admittedly hinders the detection of Blazhko modulation, the authors were able to detect light-curve modulation in about 13\,\% of first overtone RR~Lyrae (RRc) stars, and 48\,--\,72\,\% of fundamental-mode RR~Lyrae (RRab) stars. The approximately full-year coverage of the CVZs may help study Blazhko RR~Lyrae stars in these celestial regions. The authors concluded that space-based photometry is the most effective way of detecting and investigating additional pulsation modes in RR~Lyrae stars. The authors classify these features into three categories, based on their relative locations in the periodogram relative to the main pulsation frequency and its harmonics. They point out that the occurrence rates of the additional modes are the highest around the middle of the instability strip, and they decrease towards both edges.

\fullcite{2023MNRAS.521..443B} performed another comprehensive study of TESS RR~Lyrae stars. They investigated 633 RRc pulsators based on data from the first two years of observations by the space telescope. This study focused on additional frequencies, light-curve modulation, and long-term, irregular phase shifts or variations. The authors found that that the incidence rate of the Blazhko effect is 10.7\,\% among TESS RRc stars. Note that the same remark applies here that was mentioned for \fullcite{2022ApJS..258....8M} above, therefore, this value should be considered as a lower estimate. Many cases of multi-periodic modulation were also found. The authors followed year-time-scale phase variations of the dominant pulsation mode in the CVZs, and found that the phases of certain RRc stars do not actually jump, as was supposed based on ground-based seasonal observations, but they always change smoothly, in accordance with previous long-term space observations by \textit{Kepler} \combocitep{2015MNRAS.447.2348M}{2017MNRAS.465L...1S}. The phase fluctuations are correlated with the presence and number of additional modes. \fullcite{2023MNRAS.521..443B} also identified a new group of additional modes, possibly caused by nonradial $\ell = 10$ modes.

\subsubsection{Cepheid pulsators} 

\begin{figure}[]
\includegraphics[width=1.0\textwidth]{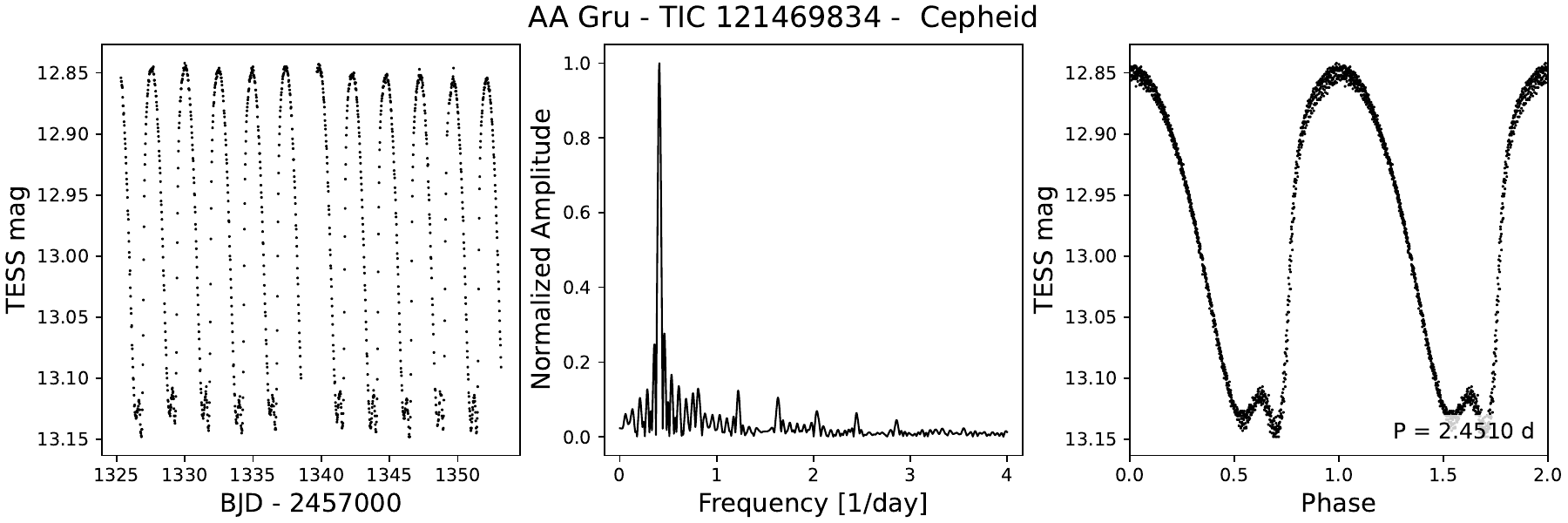}
\caption{{Representative TESS photometric data of a Cepheid star, TIC~121469834 (AA~Gru) \fullcitep{2021ApJS..253...11P}. The structure of the figure is the same as of Fig.~\ref{fig:EB}.} We used the light curve data published by \fullcite{2021ApJS..253...11P}. \label{fig:ceph}}
\end{figure}

Cepheids are a stellar class of large-amplitude radial pulsators, with characteristic periods in the 1\,--\,100\,d range. They were named after the prototype variable, $\delta$~Cephei, but later turned out that the Cepheid pulsator group consists of two main subclasses. Therefore, $\delta$~Cepheids are now considered a subgroup of Cepheids, called classical Cepheids, or type I, or population I Cepheids. The other main subgroup is called type II Cepheids which consist of BL~Her, W~Vir and RV~Tau type variables, corresponding respectively to increasing periods and different evolutionary phases. Classical Cepheids are young, massive supergiant stars that evolved from the main sequence, with masses in the 4\,--\,9\,$M_\textrm{Sun}$ range. Type II Cepheids, on the other hand, are old, low-mass stars typically in the 0.5\,--\,0.6\,$M_\textrm{Sun}$ range, burning He in a shell around their cores, and have highly inflated atmospheres and consequently high luminosities. A representative TESS light curve of a Cepheid star, TIC~121469834 (AA~Gru) is shown in Fig.~\ref{fig:ceph}.

\unskip

Cepheids are among the most important important distance indicators on the galactic and extragalactic scales. This is due to their high luminosity and the fact that they obey strict period--luminosity relations. Since the pulsation period can be easily measured with high precision even from low S/N light curves, the luminosity and consequently the distance modulus can be derived. Their brightness and characteristic, high-amplitude light variations help their identification even in distant galaxies.

Apart from this simple picture, Cepheids sometimes pulsate in multiple radial modes simultaneously, and their light curve shape can change cycle-to-cycle both irregularly or in a periodic, modulated manner. In addition, many Cepheids exhibit additional nonradial pulsation modes. To study these subtle phenomena, precise space photometry is necessary. With the almost full sky coverage and appropriate magnitude range, TESS has the potential to observe all Cepheids in the Milky Way and many more beyond our own galaxy.

The first results of TESS on 26 Galactic and Magellanic System Cepheids from Sectors~1\,--\,5 were reported by \fullcite{2021ApJS..253...11P}. The main goal of this study was to benchmark the performance of TESS in this class of variable stars. The high amplitude of the brightness variations (on the order of 1 mag) coupled with the long pulsation periods pose a challenge for the TESS photometry, since the instrument was not designed for observing such kind of brightness changes. Moreover, Cepheids tend to be situated in dense stellar fields, which, due to the low resolution of the TESS cameras, often result in strong light contamination by nearby, also potentially variable objects. Some Cepheids even saturate the photometer. Therefore, the authors developed different custom methods to optimally extract Cepheid light curves from the TESS FFI, while 2-min cadence data were also used. \fullcite{2021ApJS..253...11P} identified nonradial additional modes in several targets, and showed strong cycle-to-cycle light curve variations characteristic of W~Vir stars. The authors also emphasised that modelling unprecedentedly precise TESS light curves can potentially result in useful constraints for pulsation models. They concluded that because of the low number of the resolved objects and the length of the observations, TESS cannot compete with the OGLE Survey \fullcitep{2015AcA....65....1U} in the study of the Magellanic System. However, TESS is able to recover fine details in the light curves of certain stars in the outskirts of those galaxies.

For the above reasons, TESS data might be useful addition for Cepheid research, but this space telescope is not quite ideal tool for such variables. {Also, the longest-period Cepheids are mostly outside of the capabilities of TESS.}

\subsubsection{$\delta$~Scuti and $\gamma$~Doradus variables} 

\begin{figure}[]
\includegraphics[width=1.0\textwidth]{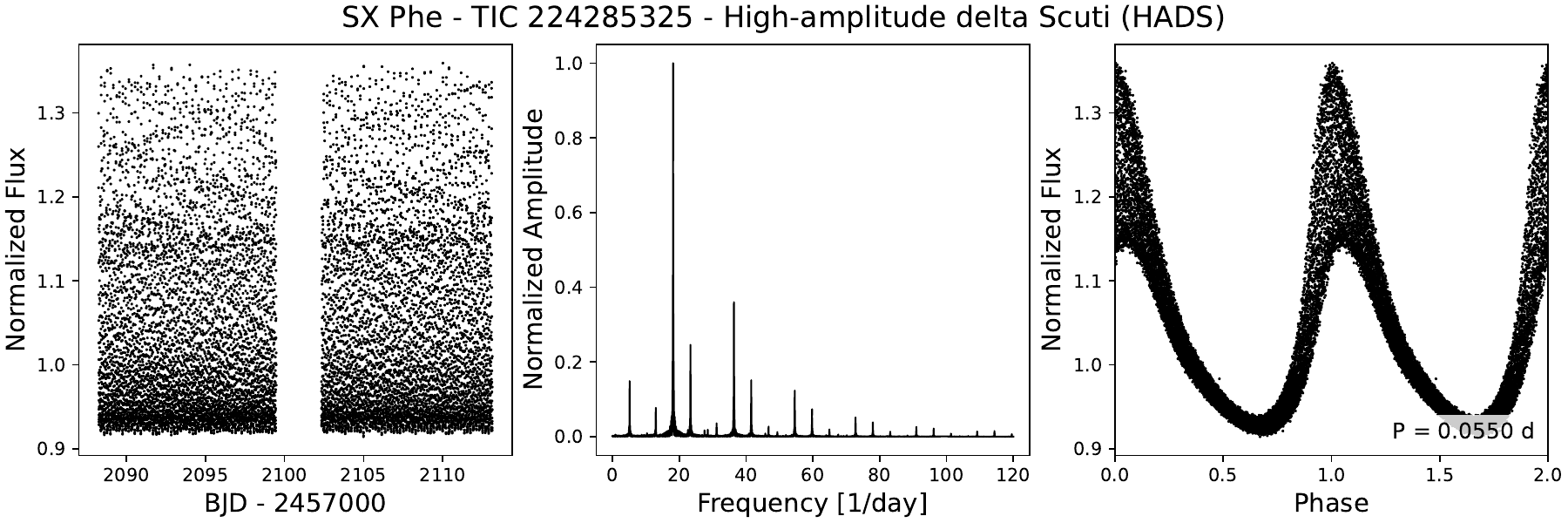}
\caption{{A representative one sector long TESS light curve of the high-amplitude $\delta$~Scuti star, TIC~224285325 (SX~Phe) \fullcitep{2019MNRAS.490.4040A}. The structure and data source of the figure is the same as of Fig.~\ref{fig:EB}.}\label{fig:dsct}}
\end{figure}

\begin{figure}[]
\includegraphics[width=1.0\textwidth]{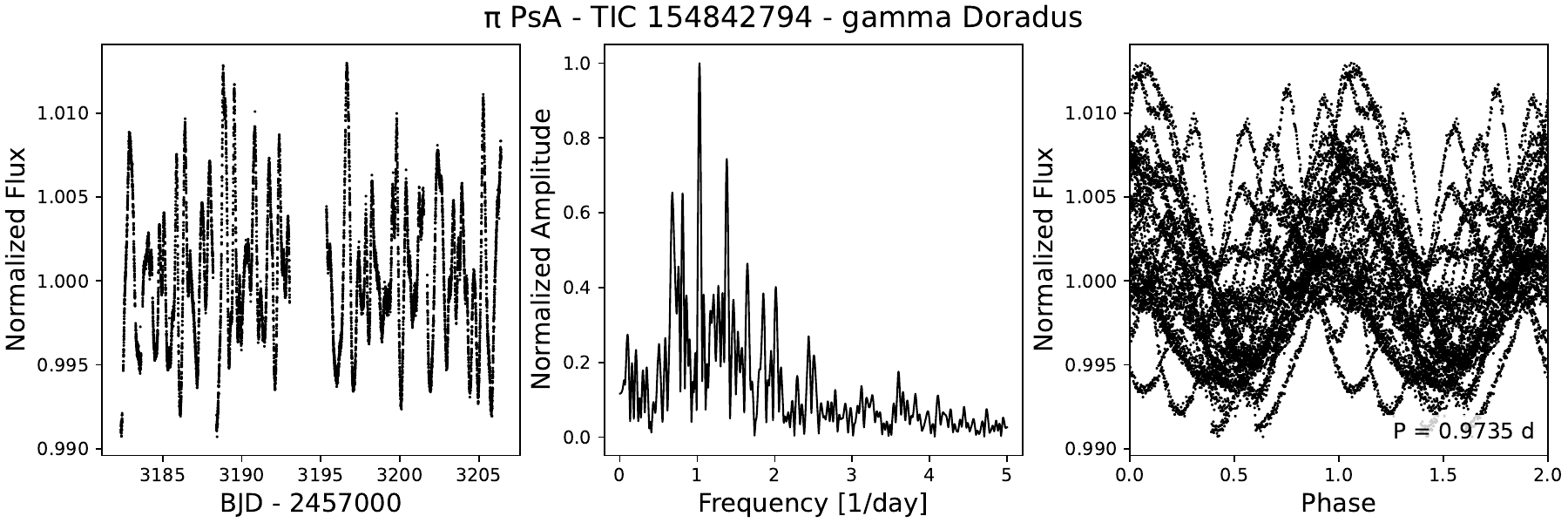}
\caption{{A representative one sector long TESS light curve of the $\gamma$~Doradus pulsator, TIC~154842794 ($\pi$~PsA) \fullcitep{2019MNRAS.490.4040A}. The structure and data source of the figure is the same as of Fig.~\ref{fig:EB}.} The spectrum in the middle panel demonstrates the low frequency resolution that data from a single TESS sector provides in the low-frequency range below 5\,d$^{-1}$, where the pulsation signal of these stars is expressed. Therefore, consecutive multi-sector data, mainly from the CVZs are preferred for $\gamma$~Doradus pulsators. \label{fig:gdor}}
\end{figure}

$\delta$~Scuti and $\gamma$~Doradus stars are intermediate-mass A and F spectral-type stars in a pre-main-sequence, main-sequence, or post-main-sequence evolutionary phase. They are situated on or near the main sequence in the HRD, where it intersects the classical instability strip. The pulsations in $\delta$~Scuti stars are excited by the opacity mechanism operating in the partial ionisation zone of \hbox{He II}, and their restoring force is pressure. These low-order radial and nonradial \textit{p}-modes have frequencies typically higher than 4\,d$^{-1}$. {Several $\delta$~Scuti variables show small amplitude ligth variations, less than 0.01\,mag. However, the so-called High Amplitude Delta Scuti stars (HADS) are characterised by much larger light variations, above 0.3\,mag.} The pulsations in $\gamma$~Doradus stars are excited by the convective flux blocking mechanism \fullcitep{2000ApJ...542L..57G}, and their restoring force is buoyancy. These low-order non-radial \textit{g}-modes have observed frequencies typically below 4\,d$^{-1}$ {and amplitudes of 0.1\,mag}. The instability regions of the $\delta$~Scuti and $\gamma$~Doradus stars overlap; therefore, hybrid pulsators that pulsate simultaneously in both regimes exist. These are usually identified as pulsating A- or F-stars with light variations in both the low- and high-frequency domains, that is, frequencies well below and above 4\,d$^{-1}$. The asteroseismic study of the interior of these pulsators helps us to understand their evolution. Some of the most important open questions are mixing processes, internal rotation, and angular momentum transfer inside these stars. Representative TESS light curves of the high-amplitude $\delta$~Scuti star, TIC~224285325 (SX~Phe), and $\gamma$~Doradus pulsator, TIC~154842794 ($\pi$~PsA) are shown in Figs.~\ref{fig:dsct} and \ref{fig:gdor}, respectively.

\unskip

\unskip

Due to their relatively short light-variation time scale, the 27\,d-long TESS sectors are well suited for studying $\delta$~Scuti pulsations, although the highest frequencies require 2-min short-cadence data. However, at the same time, due to the strongly multiperiodic nature of the $\gamma$~Doradus pulsations, considering their low frequencies, the frequency resolution of the data obtained during a single TESS sector is often insufficient for properly distinguish the unique pulsation modes of these stars, which is a concern for characterising internal rotation profiles and detailed seismic modelling. Selecting $\gamma$~Doradus targets from or near the CVZs, on which light curves from several consecutive observing sectors are available, is strongly recommended.

Thousands of $\delta$~ Scuti and $\gamma$~ Doradus pulsators were known before the TESS observations started, mainly due to the extensive data from the \textit{Kepler} space telescope (e.g. \combocite{2011MNRAS.417..591B}{2011MNRAS.415.3531B}).

\fullcite{2019MNRAS.490.4040A} performed a study based on the first two sectors of the TESS data, on how feasible TESS is to identify and study $\delta$~Scuti and $\gamma$~Doradus variable stars. They used a sample of 117 A- and F-type pulsating stars observed at 2-minute cadence by TESS. They found that the main advantages ot TESS over \textit{Kepler} in investigating $\delta$~Scuti and $\gamma$~Doradus pulsators are the near-full-sky coverage, more homogeneous sample, which includes hotter and younger $\delta$~Scuti stars, and also pre-main-sequence and metal-poor pulsators.

Thanks to the substantial sky coverage, and consequently the large sample of relatively bright targets, TESS provides many $\delta$~Scuti and $\gamma$~Doradus targets for ground-based spectroscopic follow-up studies even with smaller, 1-m class telescopes.

\subsubsection{roAp stars} 

\begin{figure}[]
\includegraphics[width=1.0\textwidth]{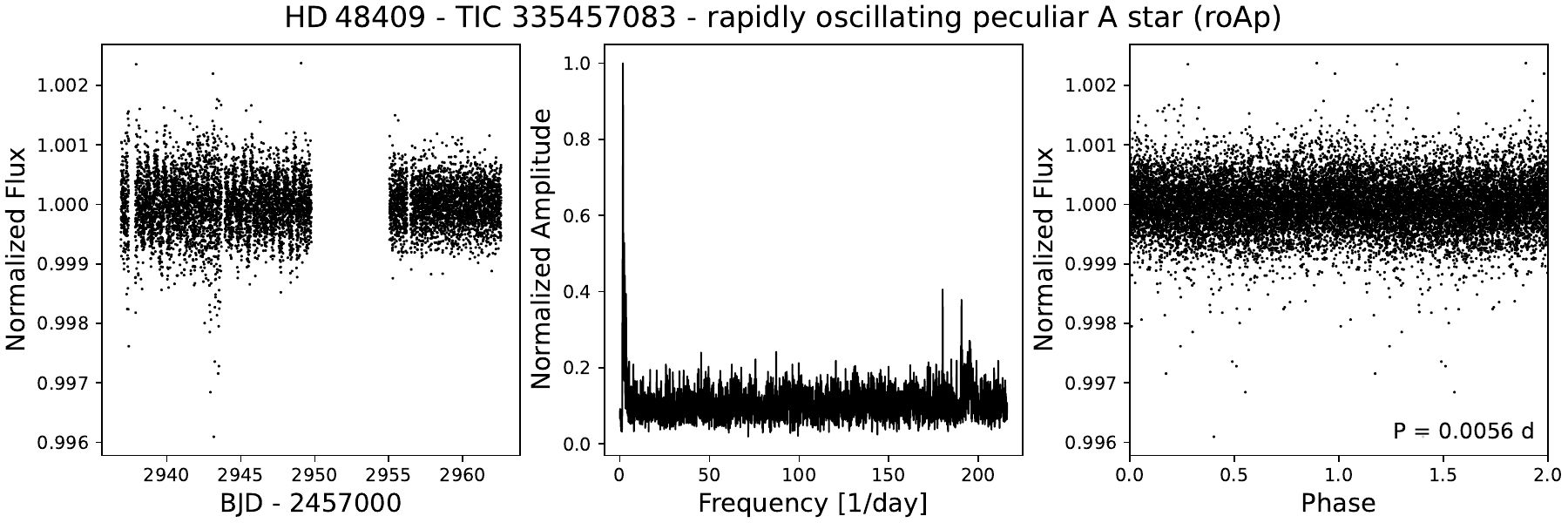}
\caption{{A representative one sector long TESS light curve of the roAp star, TIC~335457083 (HD~48409) \fullcitep{2024MNRAS.527.9548H}. The structure and data source of the figure is the same as of Fig.~\ref{fig:EB}.} The light fluctuation seen in the first half of the observations is due to rotational modulation. This signal dominates the low end of the frequency spectrum in the middle panel. The two Fourier peaks of the oscillation signal are observable between 150 and 200\,d$^{-1}$. The oscillations are very weak, and are suppressed by the rotation signal, therefore, the phased oscillation light curve in the right-hand panel does not quite show anything. \label{fig:roap}}
\end{figure}

The abbreviation roAp stands for rapidly oscillating Ap stars. Ap stars are chemically peculiar A spectral-type stars, which have a stronger magnetic field and are generally slower rotators than their classical A counterparts. These circumstances help to develop chemically overabundant patches of e.g., strontium, chromium, and europium on the stellar surface, aligned with the approximately dipole magnetic field. The magnetic axis is often oblique relative to the axis of rotation. roAp stars are a rare and oscillating subtype of Ap stars. They were discovered by \fullcite{1978IBVS.1436....1K}. Their high-overtone, low-degree, nonradial pressure modes exhibit periods in the 4\,--\,24 min range \fullcitep{2024arXiv241001715B} {and show small photometric amplitudes, around 0.1\,--\,8\,mmag}. In the HRD, roAp stars are situated at the bottom of the classical instability strip, on the main sequence, intertwined with the $\delta$~Scuti and $\gamma$~Doradus variables. {However, the stellar properties, especially the strong magnetic field}, that favour roAp pulsations, apparently suppress other, longer-period oscillations, and only one hybrid roAp--$\delta$~Scuti \fullcitep{2020MNRAS.498.4272M}, while no hybrid roAp--$\gamma$~Doradus pulsator is known. Depending on the physical conditions, the axis of the oscillation is aligned somewhere between the rotation and magnetic axes, which leads to rotational modulation of the pulsations. A representative TESS light curve of a roAp star, TIC~335457083 (HD~48409) is shown in Fig.~\ref{fig:roap}.

\unskip

TESS is quite effective at observing roAp stars, although utilisation of the 2-min cadence mode is necessary, considering the short oscillation periods. The almost full-sky coverage and sufficiently long observing runs of the individual TESS sectors provide superior time series data compared to ground-based and even to most of the previous space-based roAp observing campaigns. However, supplementary ground-based spectroscopic observations are still indispensable in studying these stars.

\fullcite{2021MNRAS.506.1073H} and \fullcite{2024MNRAS.527.9548H} conducted a comprehensive study of roAp stars observed in Cycle 1 (first year, southern hemisphere) and Cycle 2 (second year, northern hemisphere) of the TESS mission. The two studies provided homogeneous and comprehensive data on 112 confirmed roAp pulsators, of which 19 are new discoveries, while the rest had been known before the TESS observations. The TESS sample includes nearly all roAp stars known to date.

\fullcite{2021MNRAS.506.1073H} and \fullcite{2024MNRAS.527.9548H} identified many new pulsation frequencies in many roAp stars, several with multiperiodic pulsations, and concluded that frequency variability is a common feature of these objects. They found that from the sample of 112 roAp pulsators, the light curve of 71 (63\%) shows signs of rotational modulation. Moreover, \fullcite{2021MNRAS.506.1073H} and \fullcite{2024MNRAS.527.9548H} identified the longest pulsation-period and the shortest rotation-period roAp stars. They also found a new instance, TIC 402546736 ($\alpha$ Cir, HD 128898), where the pulsation modes interact nonlinearly. This is only the second known such star. Furthermore, the authors of these two studies identified several roAp pulsators with rotationally split multiplets that imply different mode geometries for the same degree modes in the same star, which provides a conundrum in applying the oblique pulsator model to the roAp stars.

We can conclude that the comprehensive and homogeneous TESS data available for the large majority of the known roAp stars represent a significant advancement in the study of this field and, complemented with the new data obtained from the still ongoing TESS observations, present important material for further studies on these objects.

\subsubsection{$\beta$~Cep and slowly pulsating B-stars} 

{The groups of $\beta$~Cep and slowly pulsating B (SPB) stars are the hottest main-sequence pulsators, of O and B spectral type. Their oscillations are excited by the $\kappa$-mechanism (see e.g. \combocite{1980tsp..book.....C}{1989nos..book.....U}), acting on the opacity bump of the iron-group elements. $\beta$~Cep stars pulsate in low-order \textit{p}- or \textit{g}-modes, with periods in the 0.1\,--\,0.3\,d range \fullcitep{2010aste.book.....A} and small amplitudes, typically only a few hundredths of a magnitude, with the highest-amplitude ones at around 0.3 mag. SPB stars pulsate in non-radial, high-radial-order \textit{g}-modes, with periods in the 0.5\,--\,3\,d range, and similarly low amplitudes than the $\beta$~Cep variables. Their instability domains overlap in the HRD, and hybrid $\beta$~Cep--SPB pulsators exist.}

{The known sample of identified $\beta$~Cep and SPB stars with available time-series data appropriate for asteroseismic studies is very limited compared to e.g. their lower-mass counterparts of known $\delta$~Sct and $\gamma$~Dor pulsators. Asteroseismology of $\beta$~Cep and SPB stars offers the opportunity of determining the interior structure, such as the size of the convective core,  overshooting, internal chemical mixing, rotation profile and internal magnetic field structure of these intermediate- and high-mass stars \fullcitep{2023ApJS..265...33S}.}

{TESS hugely contributed to the study of $\beta$~Cep stars. Beforehand, asteroseismic analysis was performed on only about a dozen such stars, due to the lack of sufficiently precise ground-based photometric and spectroscopic, and space-based photometric time series data. Altogether, on the order of  a hundred $\beta$~Cep stars were known from ground-based observations before the space photometry era \combocitep{2005ApJS..158..193S}{2005AcA....55..219P}.}

\begin{figure}[]
\includegraphics[width=1.0\textwidth]{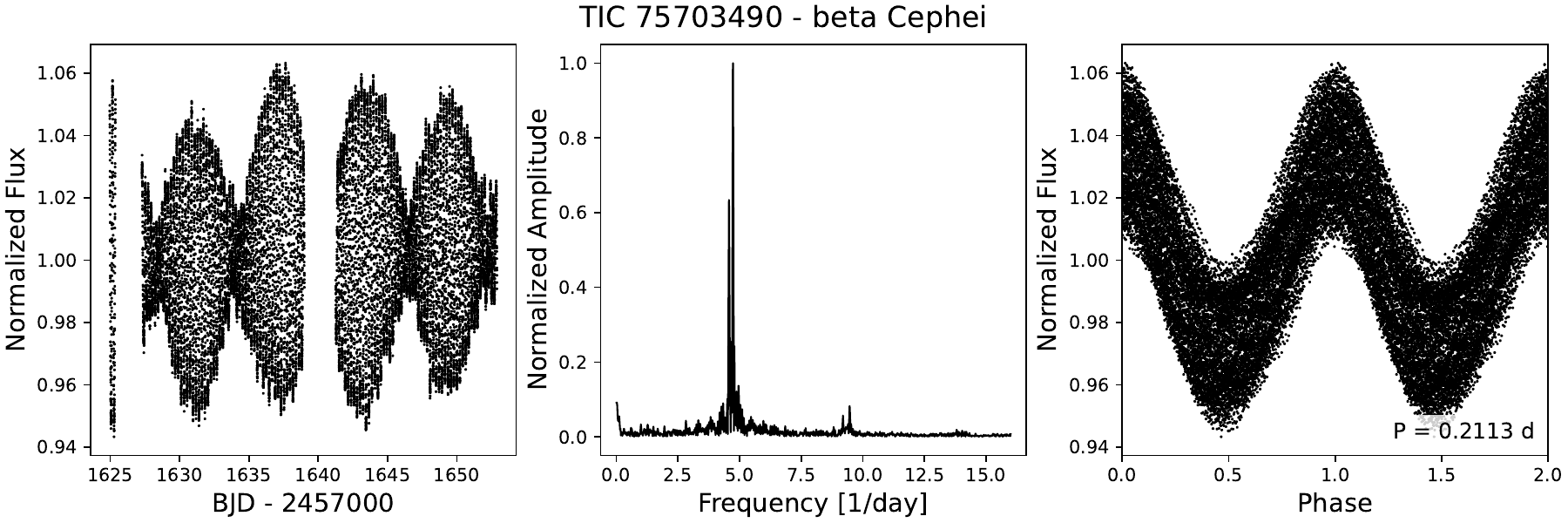}
\caption{{A representative one sector long TESS light curve of the $\beta$~Cephei star, TIC~75703490 \fullcitep{2025A&A...698A.253F}. The structure and data source of the figure is the same as of Fig.~\ref{fig:EB}.}\label{fig:bcep}}
\end{figure}

{Examining the first two cycles of 30-min-cadence TESS observations, \fullcite{2024ApJS..272...25E} identified 78 $\beta$~Cep pulsators in eclipsing binaries, of which 59 were new discoveries. Combining the variable classification of OBAF-type stars of the DR3 catalogue of the Gaia mission \fullcitep{2023A&A...674A..36G} with TESS photometry from cycles 1 and 2, \fullcite{2024A&A...688A..93H} confirmed the $\beta$~Cep classification of 222 targets. \fullcite{2025A&A...698A.253F} revisited this 222 object sample, based on the full available TESS light curve data to date. Combining TESS and Gaia data, \fullcite{2025A&A...698A.253F} successfully identified the mode degrees for 148 stars in their sample. The majority pulsate in a dominant dipole non-radial mode. They also calculated envelope rotation, spin parameter, and the level of differential envelope-to-surface rotation from the observed frequency splittings of the non-radial modes. \fullcite{2025A&A...698A.253F} found that the ratio of envelope-to-surface rotation can be as high as 3. Based on grid modelling, \fullcite{2025A&A...698A.253F} provide mass, convective core mass, and age distributions for 119 $\beta$~Cep pulsators. Hence, thanks to TESS photometric data, the sample of $\beta$~Cep stars studied by asteroseismic means increased about tenfold. A representative TESS $\beta$~Cep light curve of TIC~75703490 is shown in Fig.~\ref{fig:bcep}.}

\unskip

\begin{figure}[]
\includegraphics[width=1.0\textwidth]{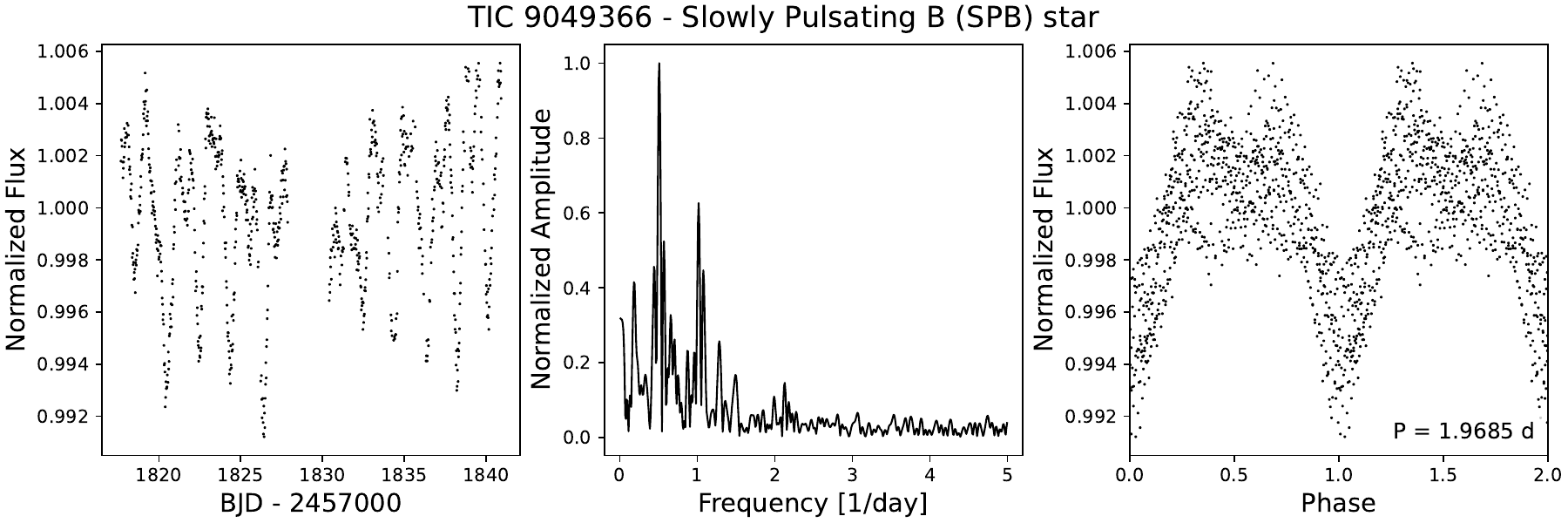}
\caption{{A representative one sector long TESS light curve of the SPB star, TIC~9049366 (BD+55 884) \fullcitep{2023ApJS..268...16S}. The structure and data source of the figure is the same as of Fig.~\ref{fig:EB}. All the frequencies of the detected light variations are below 3\,d$^{-1}$.}\label{fig:SPB}}
\end{figure}

{TESS also provides significant contribution to the study of SPB variables. Combining TESS, Gaia and LAMOST \fullcitep{2012RAA....12.1197C} data, \fullcite{2023ApJS..265...33S} and \fullcite{2023ApJS..268...16S} identified 87 and 286 new SPB stars, respectively. \fullcite{2023ApJS..268...16S} found that, apart from the HRD, the period--temperature (P--T) and period--luminosity (P--L) diagrams are also useful tools to identify both $\beta$~Cep and SPB variables. A representative TESS SPB light curve of TIC~9049366 is shown in Fig.~\ref{fig:SPB}.}

\unskip

{Unfortunately, due to the relatively long pulsation periods of SPB stars, compared to the typical length of TESS data strings of 27-d sectors, the frequency resolution of these data sets are often insufficient for detailed asteroseismic inferences. The most promising targets are those observed for about a full year in the CVZ. We expect more detailed case studies and results in the future, based on TESS data of such SPB stars.}

\subsubsection{Solar-like oscillations} 

Solar-like oscillations are stellar oscillations excited in the same manner as those in the Sun, by turbulent convection in the outer layers. The oscillations consist of standing pressure modes and mixed modes, excited over a range of frequencies. Surface convection damps these modes, and each mode can be well approximated in the frequency space by a Lorentzian curve, whose width corresponds to the mode's lifetime: the Lorentzian is broader if the decay is faster.

\begin{figure}[]
\includegraphics[width=1.0\textwidth]{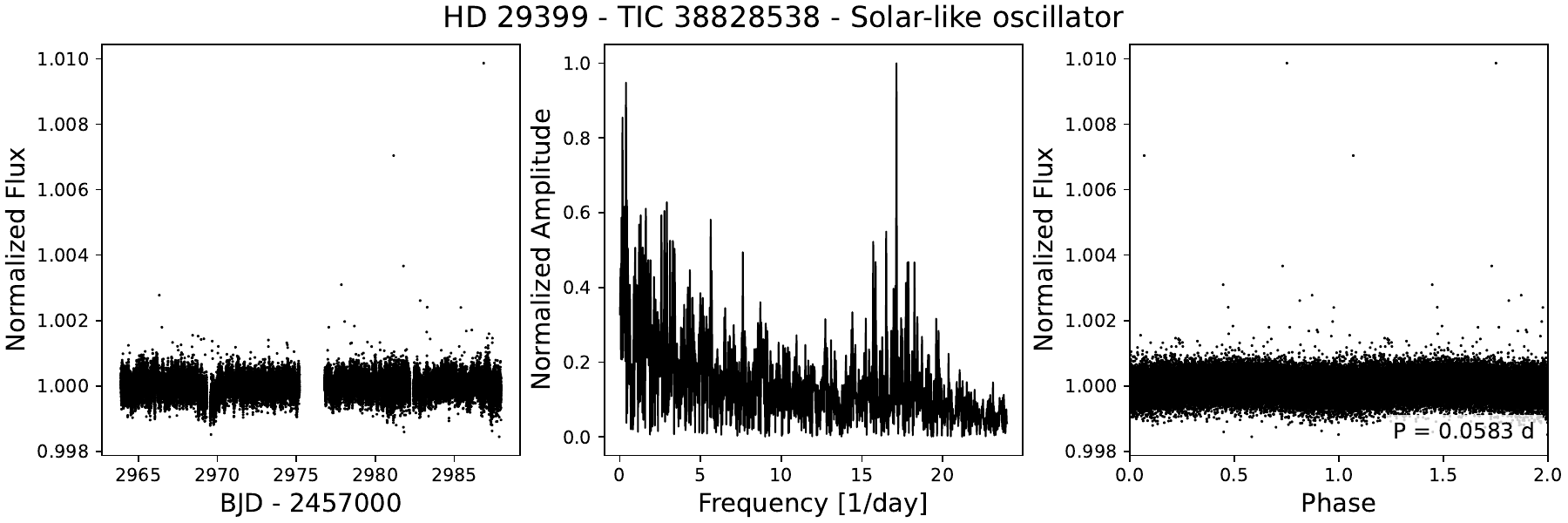}
\caption{{A representative one sector long TESS light curve of the solar-like variable star, TIC~38828538 (HD~29399) \fullcitep{2024ApJS..271...17Z}. The structure and data source of the figure is the same as of Fig.~\ref{fig:EB}.} The upwards strongly outlying data points in the light curve are not measurement errors, but they represent stellar flares, caused by magnetic activity. Due to the low-amplitude, but strongly multiperiodic oscillation signal, represented in the middle panel by many Fourier peaks in the 15\,--\,20\,d$^{-1}$ frequency range, the light curve in the right-hand panel, phased by the dominant oscillation period does not quite show the oscillation itself. \label{fig:solarlike}}
\end{figure}

We expect that all stars possessing surface convection zones are likely to show solar-like oscillations, that is, red giant and subgiant stars, and cool main-sequence stars. The revolutionary observations of photometric space telescopes allowed for the detection of solar-like oscillations in many stars because these oscillations have small amplitudes. The observations of solar-like oscillations also help to improve determinations of exoplanetary masses and radii, since by studying them, we are able to precisely determine the masses and radii of the planet-hosting stars. A representative TESS light curve of a solar-like oscillator, TIC~38828538 (HD~29399) is shown in Fig.~\ref{fig:solarlike}. Solar-like oscillations are detected also in binary systems such as 12~Bo\"otis \fullcitep{2022MNRAS.516.3709B}.

\unskip

From an asteroseismological point of view, observations with \textit{Kepler} revealed in the case of red giants that the observed oscillations can be used to distinguish between objects that burn helium in their cores and those still show shell hydrogen burning \fullcitep{2011Natur.471..608B}.

Considering TESS, 20 and 120\,s cadence mode observations are suitable for observations of solar-like oscillations. \fullcite{2023A&A...669A..67H} reported the investigation of sectors 1\,--\,46, and compiled a catalogue of 4177 solar-like oscillators. They also determined the two most fundamental asteroseismic indicators of this class of variable stars, the large frequency separations ($\Delta\nu$) and the frequencies of maximum power ($\nu_{\mathrm{max}}$) for 98\% of these stars. These targets can be found on the red giant branch, the subgiant regime, and towards the main sequence. Knowing the large frequency separation allows us to infer the star's density, and if its effective temperature is also known, we can determine its mass and radius as well. Knowing the frequency corresponding to the maximum pulsation energy allows us to infer the star's surface gravity. By combining this with information from the large frequency separation, we can also determine its mass and radius, thus the evolutionary status of the object can completely be inferred.

\fullcite{2024ApJS..271...17Z} published a catalogue of solar-like oscillators. They investigated sectors 1\,--\,60, and provided a catalogue of 8651 solar-like variables, including 2173 new discoveries. They also estimated asteroseismic masses, radii, and $\mathrm{log}g$ for a subset of 7173 stars.

Many exoplanet host stars show solar-like oscillations, therefore, this area of study can also help promoting the main programme of TESS, by characterising the discovered exoplanetary systems.

\subsubsection{Pulsating white dwarfs} 

\begin{figure}[]
\includegraphics[width=1.0\textwidth]{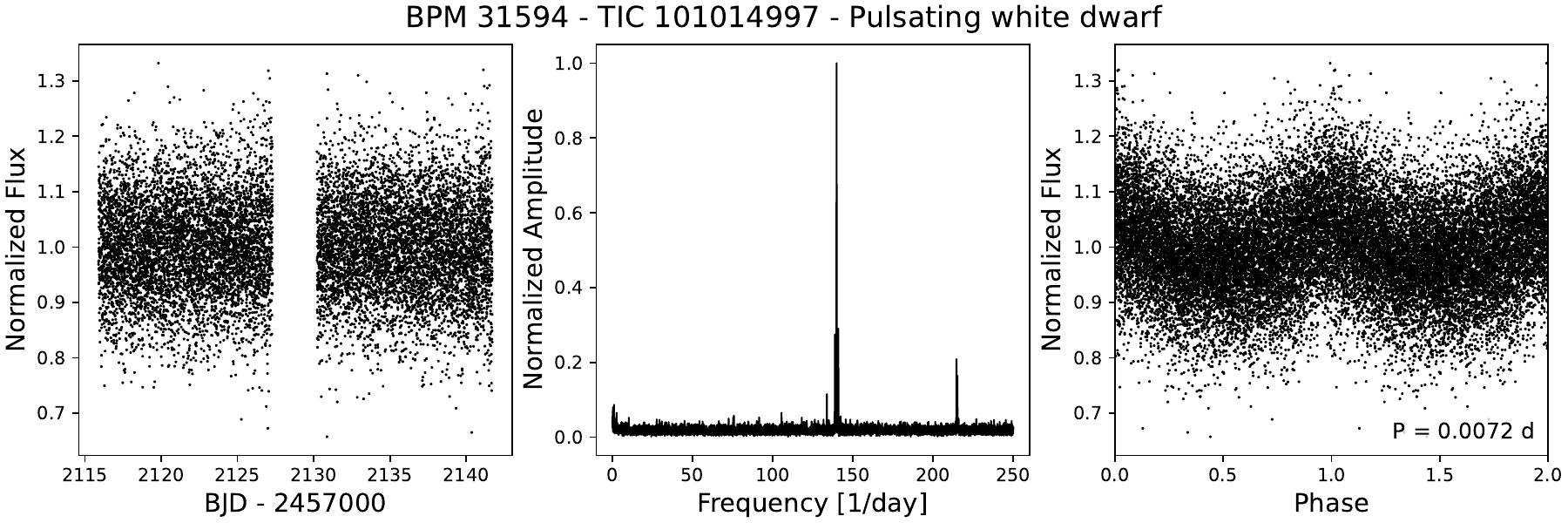}
\caption{{A representative one sector long TESS light curve of the pulsating white dwarf star, TIC~101014997 (BPM~31594) \fullcitep{2020A&A...638A..82B}. The structure and data source of the figure is the same as of Fig.~\ref{fig:EB}.}\label{fig:pwd}}
\end{figure}

White dwarfs represent the final evolutionary stage of stars with low-to-intermediate initial mass,
comprising about 98\% of all stars, including our Sun (see the e.g. the review of \fullcite{2008ARA&A..46..157W} and references therein). The astrophysical significance of white dwarfs is abundant. They are used to study the early star formation history, the evolution, age, and structure of the Galactic disk. Some white dwarf stars are among the oldest known objects in the Universe. Most of the planet-host stars will also eventually become white dwarfs. White dwarfs also can be used as cosmic laboratories to study matter under extreme physical conditions, e.g. the properties of high-density plasmas, degenerate or even crystallised stellar matter. White dwarfs also play an important role in understanding stellar evolution, as their properties, such as cooling times, the chemical composition of their core, and their mass distribution are key parameters to constrain the stellar evolution theory, e.g. to understand the mass loss processes on the asymptotic giant branch of the HRD (see \fullcite{2019A&ARv..27....7C} and references therein). A representative TESS light curve of a pulsating white dwarf star, TIC~101014997 (BPM~31594) is shown in Fig.~\ref{fig:pwd}.

\unskip

Some of the white dwarfs show low-amplitude ($A \sim 0.1\%$) and short-period light variations due to oscillations. We
can find them in well-defined regions of the HRD, and they are classified into three main
groups: GW\,Vir, V777\,Her (DBV) and ZZ\,Ceti (DAV) stars. GW Vir (pre)white dwarfs are the hottest stars with effective temperatures between about 80\,000\,--\,180\,000\,K and hydrogen-deficient atmospheres, while DBV and DAV stars are much cooler with effective temperatures of about 22\,000\,--\,32\,000\,K and 10\,500\,--\,13\,000\,K and neutral helium and hydrogen-dominated atmospheres, respectively. ZZ Ceti stars constitute the most populated class; about 80\% of all known pulsating white dwarfs belong to this group. The brightness variations of GW Vir, DBV, and DAV stars are caused by non-radial pulsations driven by different physical mechanisms. GW Vir stars exhibit pulsations due to the $\kappa$-mechanism operating in the partial ionization zone of carbon and oxygen. DBVs are excited by a combination of the $\kappa$-mechanism acting in the He partial ionization zone, and the so-called convective driving mechanism. In DAV stars, the $\kappa - \gamma$ mechanism and also the convective driving mechanism are responsible for the excitation of the observed pulsations. 

\begin{figure}[]
\includegraphics[width=1.0\textwidth]{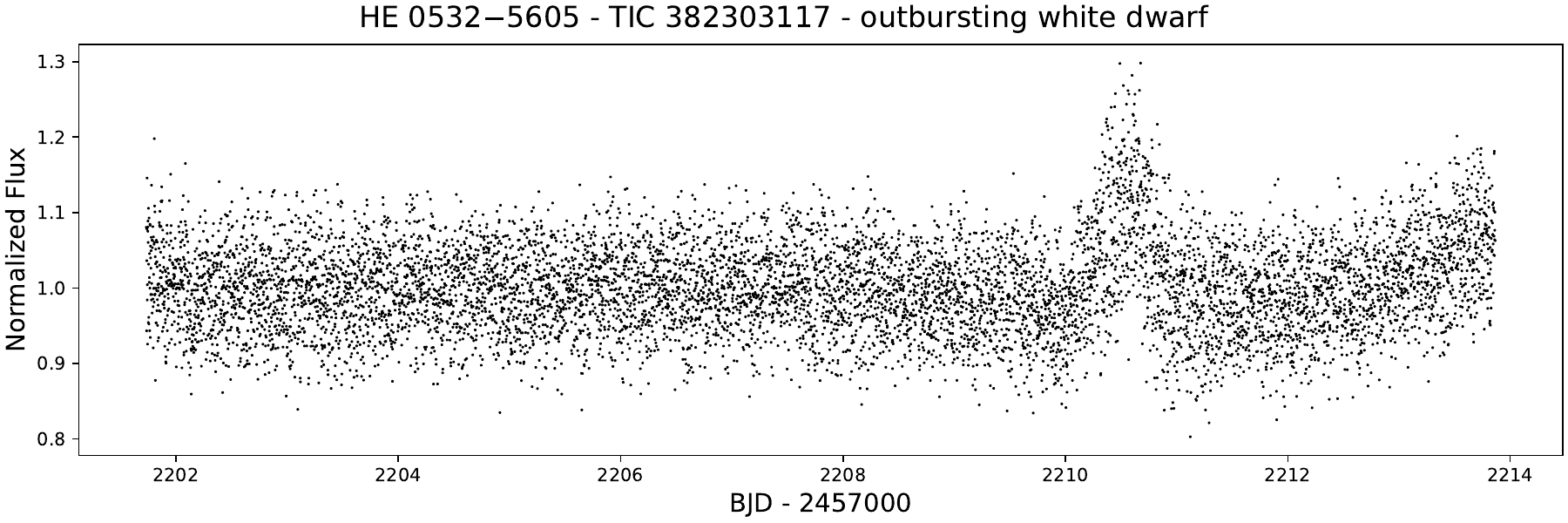}
\caption{{Outburst seen in TESS sector~1 of HE 0532-5605 \fullcitep{2020A&A...638A..82B}. The data source and processing are the same as of Fig.~\ref{fig:EB}. The outburst event appears between 2210 and 2211\,d on the scale of the horizontal axis. Another outburst was also detected in sector~33.} \label{fig:outburst}}
\end{figure}

Using TESS light curves, \fullcite{2020A&A...638A..82B} and \fullcite{2023A&A...674A.204B} detected several new pulsation frequencies in the studied pulsating white dwarf stars, which helps imposing stronger constraints for future asteroseismic investigations. They also found that the star HE\,0532-5605 is a new outbursting ZZ~Ceti. The outburst phenomenon is an increase in the stellar flux of ZZ~Ceti stars close to the red edge of the instability strip. Observations show that the average brightness of the star increases relatively quickly (in about an hour) by at least several per cent and remains in this elevated state for several hours, sometimes even for a day or more. After that, the stellar brightness decreases to the initial level; the outburst event may repeat after several days, weeks, or more, as it was found in HE~0532-5605 \fullcitep{2023A&A...674A.204B}, as shown in Fig.~\ref{fig:outburst}. The duration and occurrence of these events is irregular and unpredictable.

\unskip

Compared to \textit{Kepler}, TESS has detected a much larger number of pulsating white dwarf stars: an important TESS result considering these objects is the significant increase in the number of newly detected DAVs. \fullcite{2022MNRAS.511.1574R} discovered 74 new DAV stars analyzing the TESS data from sectors 1\,--\,39 (the first three years). They investigated 8300 known white dwarfs to find new pulsators. They extended the work by reviewing the data obtained during the fourth and fifth years of the TESS mission and presented the discovery of 32 additional DAVs in 2024 \fullcitep{2025ApJ...984..112R}.

\subsubsection{Pulsating hot subdwarf stars} 

\begin{figure}[]
\includegraphics[width=1.0\textwidth]{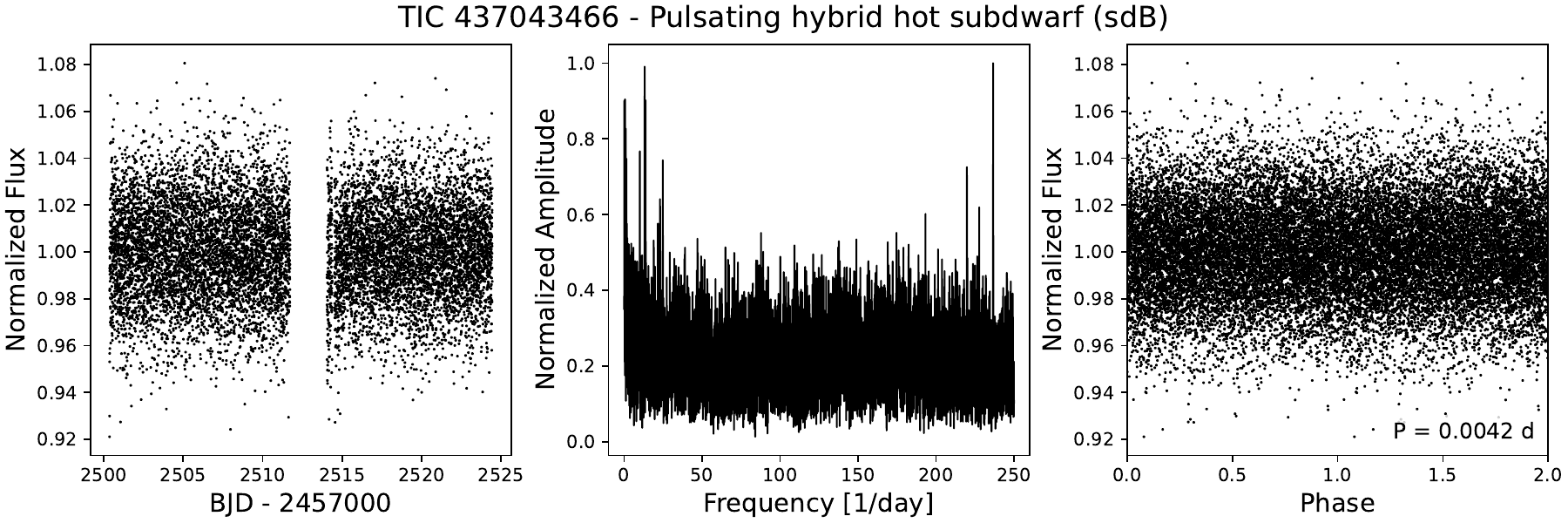}
\caption{{A representative one sector long TESS light curve of the pulsating hot subdwarf star, TIC~437043466 \fullcitep{2023A&A...669A..48B}. The structure and data source of the figure is the same as of Fig.~\ref{fig:EB}.} This is a hybrid pulsator, meaning that it shows low-frequency \textit{g}- and high-frequency \textit{p}-modes simultaneously, which appears in the Fourier spectrum in the middle panel as peaks both below 30\,d$^{-1}$ and in the 200\,--\,250\,d$^{-1}$ high-frequency range. Due to the strong multiperiodicity and since the highest-amplitude pulsation mode is not very dominant, the light variations cannot really be seen in the folded light curve of the right-hand panel. \label{fig:sdb}}
\end{figure}

Hot subdwarf B-type stars (sdB stars) are helium core burning objects with an extremely thin hydrogen envelope of less than 0.02\,\,$M_\textrm{Sun}$ (see, e.g. \fullcite{2022ApJ...933..137G}). The average mass of sdB stars is approximately 0.47\,\,$M_\textrm{Sun}$ \fullcitep{2012A&A...539A..12F}. Their surface gravity ranges from 5.2 to 6.2\,dex, and their effective temperatures are between 20\,000 and 40\,000\,K (\combocite{1994ApJ...432..351S}{2008ASPC..392...75G}). These stars are located between the main sequence and the cooling track of white dwarfs, on the so-called extreme horizontal branch (EHB; \combocite{2009ARA&A..47..211H}{2016PASP..128h2001H}) of the HRD. A representative TESS light curve of a hybrid pulsating hot subdwarf star, TIC~437043466 is shown in Fig.~\ref{fig:sdb}.

\unskip

The evolutionary history of sdB stars involves significant mass loss, mainly driven by binary interactions during the late stages of the red giant phase (\combocite{2002MNRAS.336..449H} {2003MNRAS.341..669H}). As a result, nearly the entire hydrogen-rich envelope is lost, leaving behind a helium-burning core with an envelope too thin to sustain hydrogen shell burning. The sdB stars continue to burn helium in their cores for approximately 100 million years. When the helium content in the core drops below a critical level, the star enters a phase where helium fusion continues in a shell surrounding a carbon-oxygen core. At this stage, the star evolves into an sdO star. Eventually, these stars complete their life cycle as white dwarfs (see, e.g. \fullcite{1993ApJ...419..596D}).

The existence of pulsating sdB stars was predicted by \combocite{1996ApJ...471L.103C} {1997ApJ...483L.123C}. The first pulsating sdB star was discovered by \fullcite{1997MNRAS.285..640K}. They identified EC\,14026-2647, which exhibits short-period $p$-mode pulsations. Later, \textit{g}-mode pulsators, and also hybrid \textit{p}- and \textit{g}-mode sdB variables (sdBV stars) were discovered (\combocites{2003ApJ...583L..31G}{2005MNRAS.360..737B}{2006A&A...445L..31S}). {The amplitudes of sdBV variables are typically around 0.1\% of the mean flux.} 

The study of sdBV pulsations has proven useful for precise mass estimations of sdB stars using the tools of asteroseismology. The pulsations in sdBV stars are primarily driven by the $\kappa$-mechanism, operating in the partial ionization zone of iron-group elements.

The first-light paper on TESS sdBV stars presents the result of the light curve analysis and the asteroseismic investigations on the $g$-mode pulsator EC\,21494-7018 \fullcitep{2019A&A...632A..90C}. The authors determined 20 independent modes for asteroseismology, which revealed that its mass ($0.391 \pm 0.009$\,\,$M_\textrm{Sun}$) is significantly lower than the typical mass of sdB stars. This leads to the presumption that its progenitor was a massive red giant, which has not undergone a He-core flash. This means an alternative channel for the formation of sdB stars.

A series of papers deals with the search for $p$-mode variables (\combocite{2023A&A...669A..48B} {2024A&A...686A..65B}). The first paper presents the identification of 32 sdBV, and 8 hybrid sdB pulsators in the southern ecliptic hemisphere. The second study examines the northern ecliptic hemisphere, where the authors found 35 sdBVs and 9 hybrid pulsators.   

We also have to mention two papers by \combocite{2021A&A...651A.121U}{2024A&A...684A.118U}. In the first study, the authors performed a detailed asteroseismic and spectroscopic analysis of five $g$-mode TESS sdBV stars. The second paper reports the discovery of 61 new sdBVs, which in most cases show several pulsational frequencies, extracted from the TESS photometry.

\section{Conclusion}

In the previous sections, we have covered a long journey highlighting many important scientific results based on TESS observations. We introduced objects that exhibit brightness variations due to extrinsic or intrinsic causes and highlighted some key findings for each object type based on TESS measurements. It can be stated that TESS is successfully fulfilling its mission; even from the results so far, it is clear that for every type of object we presented in this paper, we have gained new insights thanks to TESS and the dedicated team behind it.

TESS has revolutionised both the search for exoplanets, and stellar variability by providing high-precision photometric data across nearly the entire sky. Its discoveries help shape our understanding of planetary systems, stellar evolution, and the broader astrophysical landscape. With the ongoing extended missions, TESS is remaining to be a critical tool for both planetary and stellar astrophysics in the coming years.


\vspace{6pt} 


\funding{The authors acknowledge the financial support of the KKP-137523 `SeismoLab' \'Elvonal grant of the Hungarian Research, Development and Innovation Office (NKFIH). Funding for the TESS mission is provided by the NASA Explorer Program. STScI is operated by the Association of Universities for Research in Astronomy, Inc., under NASA contract NAS 5–26555.}

\conflictsofinterest{The authors declare no conflicts of interest. The funders had no role in the design of the study; in the writing of the manuscript; or in the decision to publish the results.} 

\textbf{Data Availability Statement:} No new data were created in connection with this work. The authors only used publicly available data for generating the figures. The source of these data are provided in the captions of the figures.


\begin{adjustwidth}{-\extralength}{0cm}

\reftitle{References}



\bibliography{TESS_review}

\PublishersNote{}
\end{adjustwidth}
\end{document}